\documentclass[showpacs,aps,pra,amsmath,superscriptaddress,twocolumn]{revtex4-1}

\usepackage{mathrsfs}
\usepackage{amssymb}
\usepackage{amstext}
\usepackage{mathtools}
\usepackage[pdftex]{color}
\usepackage[english]{babel}
\usepackage{graphicx}
\usepackage{bm}
\usepackage{float}
\usepackage{ulem}

\makeatletter

\newcommand{\Rmnum}[1]{\expandafter\@slowromancap\romannumeral #1@}
\makeatother

\begin{document}

\title{Quantum Vertex Model for Reversible Classical Computing}

\author{C. Chamon}

\email[Corresponding author: ]{chamon@bu.edu}

\affiliation{Physics Department, Boston University, 590 Commonwealth
  Ave., Boston, Massachusetts 02215, USA}

\author{E. R. Mucciolo}

\affiliation{Department of Physics, University of Central Florida,
  4111 Libra Drive, Orlando, Florida 32816, USA}

\author{A. E. Ruckenstein}

\affiliation{Physics Department, Boston University, 590 Commonwealth
  Ave., Boston, Massachusetts 02215, USA}

\author{Z.-C. Yang}

\affiliation{Physics Department, Boston University, 590 Commonwealth
  Ave., Boston, Massachusetts 02215, USA}

\date{\today}

\begin{abstract}

Mappings of classical computation onto statistical mechanics models
have led to remarkable successes in addressing some complex
computational problems. However, such mappings display thermodynamic
phase transitions that may prevent reaching solution even for easy
problems known to be solvable in polynomial time. Here we map
universal reversible classical computations onto a planar vertex model
that exhibits no bulk classical thermodynamic phase transition,
independent of the computational circuit.  Within our approach the
solution of the computation is encoded in the ground state of the
vertex model and its complexity is reflected in the dynamics of the
relaxation of the system to its ground state. We use thermal
annealing with and without ``learning'' to explore typical
computational problems. We also construct a mapping of the vertex
model into the Chimera architecture of the D-Wave machine, initiating
an approach to reversible classical computation based on
state-of-the-art implementations of quantum annealing.

\end{abstract}

\maketitle

\section{Introduction}
\label{sec:intro}

Throughout the past few decades, problems of computer science have
become subjects of intense interest to theoretical physicists as
paradigms of complex systems that could benefit from theoretical
approaches and insights inspired by statistical physics. These include
neural networks, Boltzmann machines and deep learning, compressed
sensing, satisfiability problems, and a host of other approaches to
data mining and machine
learning~\cite{montanari_mezard,MPZ,ganguli,pankaj}. The interest in
the constraints on computation and information processing placed by
physical laws is even older and dates to work by Landauer and
Bennett~\cite{landauer1,landauer2,bennett}. One of the holy grails at
the interface between physics and computer science is the physical
realization of a large-scale quantum computer in which the processing
of information makes use of quantum-mechanical concepts such as
superposition and entanglement~\cite{nielsen-chuang,review}. However,
building a quantum computer remains a challenging task because of the
practical difficulty associated with maintaining coherence over the
duration of the computation.

This paper aims at bringing a new class of problems to the
physics-computer science interface by introducing a two-dimensional
(2D) representation of a generic reversible classical computation, the
result of which is encoded in the ground state of a statistical
mechanics vertex model with appropriate boundary conditions. The
vertex model is defined in terms of Boolean variables (or spins
degrees of freedom) placed on the bonds or links of an anisotropic 2D
lattice with vertices representing logic gates. The corresponding gate
constraints are implemented through short-ranged one- and two-body
interactions involving the spins of the vertex (as we show, this
construction can be realized in physical programmable machines, such
as the D-Wave machine.) One direction of the lattice represents
``computational (rather than real) time'', as introduced by Feynman in
the history representation of quantum computation~\cite{feynman}, but
here used for classical reversible circuits. The two boundaries of the
lattice transverse to the ``time'' direction contain the input and
output bits of the computation. It is important to stress that we are
not limiting ourselves to forward computations with fixed inputs. More
interesting are problems in which only partial information about both
inputs and outputs is known. In that case, reaching the ground state
requires flow of information both forwards and backwards across the
lattice, processes that are naturally built into our approach.

The idea of encoding classical computation in the ground state of a
many-body spin model was introduced earlier for irreversible
computation in Ref.~\cite{Biamonte2008,crosson2010,Biamonte2012}. Here
we focus on reversible rather than irreversible computation in order
to address problems with both fixed-input and mixed-boundary
conditions on inputs and outputs, as explained above. Mapping onto a
regular 2D lattice as opposed to an arbitrary graph allows us to use
intuitive ideas from equilibrium and non-equilibrium statistical
mechanics, especially of classical and quantum phase
transitions. Also, while in Ref.~\cite{crosson2010} an error
correction scheme was required to implement fault tolerant
computation, in our approach accurate computation without error
correction is possible below moderate temperatures that scale only as
the inverse of the logarithm of the system size, a consequence of the
exponential scaling of the static correlation length with inverse
temperature (see below). 

Most importantly, the mapping proposed here defines statistical
mechanics vertex models that, irrespective of the computation they
represent, display no bulk thermodynamic transition down to zero
temperature. Thus our work emphasizes that the
dynamics of relaxation to the ground state rather than the
thermodynamics of the model is essential for understanding the
complexity of ground state computation.

The absence of a thermodynamical phase transition removes an obvious
impediment to reaching the ground state of the vertex model. For
instance, a suboptimal mapping from a computational problem into a
physical system may place the solution within a glassy phase, even in
the case of easy computational problems. The mapping of XORSAT (a
problem in P) into a diluted $p$-spin model is such an
example~\cite{Ricci-Tersenghi}. The fact that our vertex model is free
of thermodynamic transitions does not mean that the ground state can
be reached easily. This remains true even for problems with unique
solutions which are encoded by vertex models with unique ground
states.  Such problems are in the complexity class UNIQUE-SAT, which
under randomized reduction is as hard as
SAT~\cite{Valiant-Vazirani}. Hence, even in the absence of a
thermodynamic transition finding the unique ground state of vertex
models encoding problems with a single solution is a problem in
NP-complete~\cite{gavey-johnson,papadimitriou,arora-barak}. Of course,
this does not mean that one cannot benefit from speed-ups allowed by
either physics inspired heuristics or by special-purpose physical
hardware, such as quantum annealers.

This paper focuses on the study of vertex-model representations of
random circuits for which the complexity of the computation is
reflected in the concentration of TOFFOLI gates, the length of the
input and output boundaries $L$, and the depth of the circuit $W$. We
concentrate on computational problems with a single solution -- or
problems for which one can discern among an ${\cal O}(1)$ number of
solutions with a small overhead -- a class of problems that encompass
factoring of semi-primes, an important and nontrivial example that we
shall explore in a future publication.

In our discussion of dynamics we deploy thermal annealing as well as
introduce a more efficient ``annealing with learning'' protocol. The
latter translates into an algorithm for solving classical problems for
which, as expected, forward computation from a fixed input boundary
reaches solution in a time linear in the depth of the computational
circuit.  Finally, we note that reaching the ground state of the
vertex model could be accelerated by replacing classical annealing
with quantum
annealing~\cite{apolloni1989,finnila1990,kadowaki1998,farhi2001}. While
approaching computational problems through quantum annealing is left
for future investigations, the current paper includes the formal
derivation of the quantum version of the statistical mechanics model
of reversible classical computation. This provides the background for
an explicit mapping of our lattice model onto the Chimera architecture
of the D-Wave machine, a development that points to the potential
usefulness of the vertex model as a programming platform for special
purpose quantum annealers.

\section{Results}
\label{sec:Results}

\subsection{The vertex model for reversible classical computation}
\label{sec:tile-model}

\begin{figure}[t]
\centering
\includegraphics[angle=0,origin=c,width=9cm]{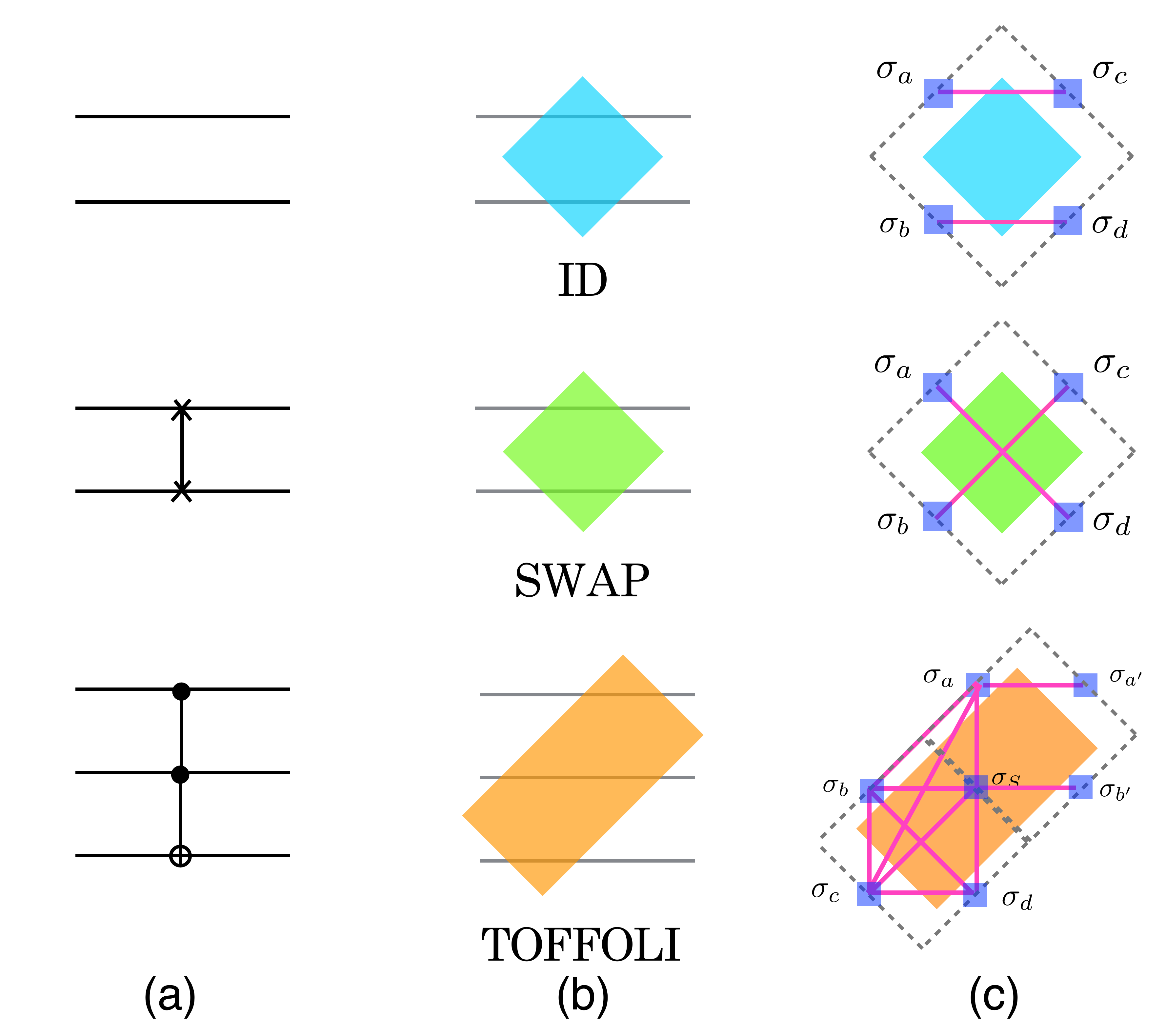}
\caption{{\bf Tile representation of reversible computational
    gates.} (a)\&(b) Elementary tiles representing the three
  computational gates for reversible circuits: ID (identity), SWAP and
  TOFFOLI. (c) construction of the Hamiltonians that encode the
  gate-satisfying states in the ground state manifolds. Spins are
  placed on the boundary of the tiles. For the TOFFOLI gate, an
  ancilla spin is placed in the center of the rectangular
  tile. Couplings needed in the Hamiltonians for the three different
  gates (tiles) are indicated by purple lines connecting two
  spins. The dashed line denotes the boundary of the tile.}
\label{fig:gate_tile_coupling}
\end{figure}

Our starting point is the fact that any Boolean function can be
implemented in terms of TOFFOLI gates, which are reversible logic
gates with three inputs and three outputs. Starting from a circuit of
TOFFOLI gates, our construction proceeds by first using SWAP gates to
repeatedly swap distant bits in the input that are acted upon by
particular gates of the circuit, until the operation of every gate is
reduced to adjacent bits. The second step is to associate tiles with
each of the gates, as shown in Fig. ~\ref{fig:gate_tile_coupling},
where one should imagine placing input and output bits at the
intersections of the tile surfaces with the horizontal lines, as
described in detail in the Methods Section.

\begin{figure}[hbt]
\centering
\includegraphics[angle=0,origin=c,width=8.5cm]{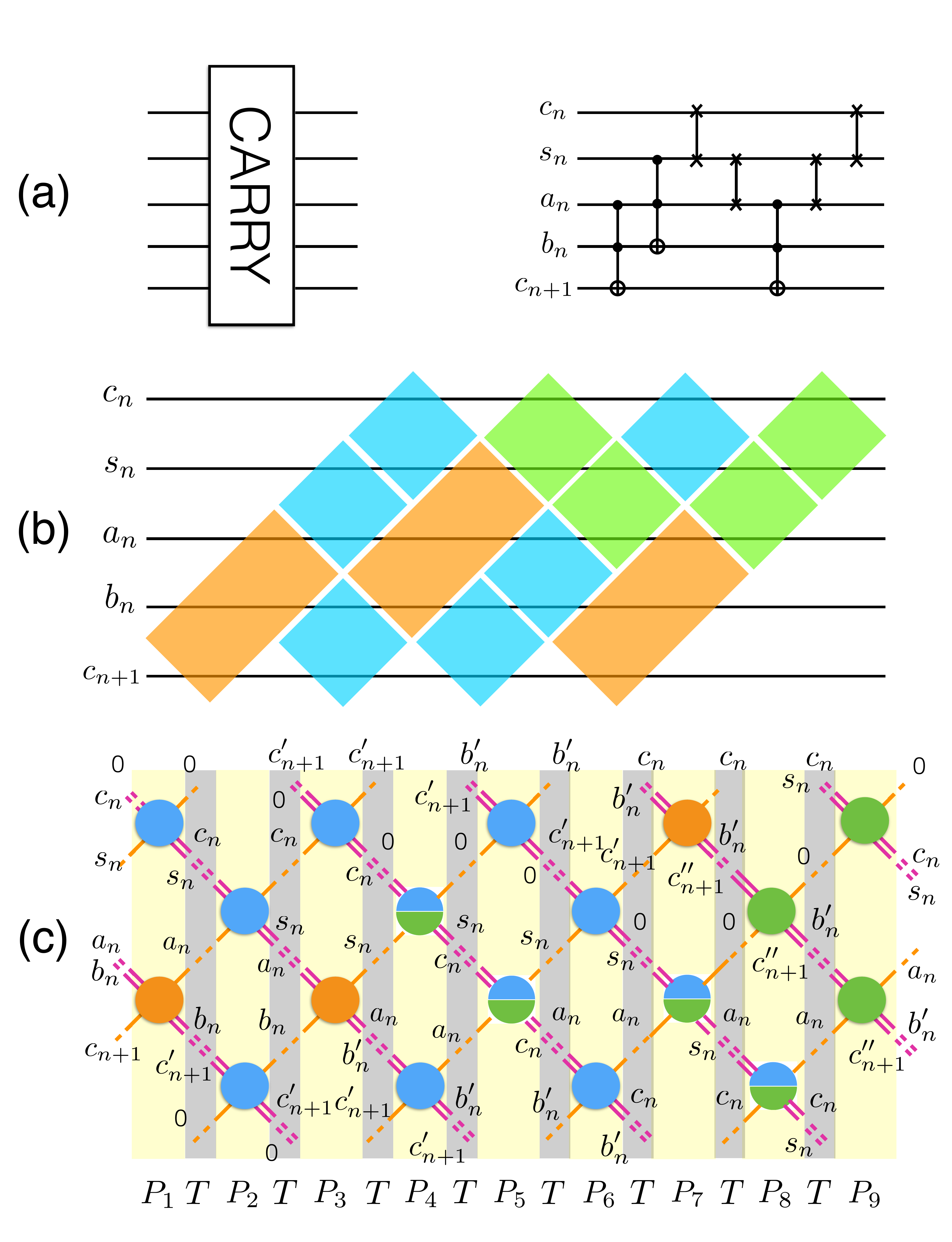}
\caption{{\bf Tile and vertex model representation of the
    ripple-carry adder.} (a) The ripple-carry adder which computes the
  carry bit that is ``rippled'' to the next bit. We add one additional
  control line $s_n$ and set it to 1 to implement the original CNOT
  gate with a TOFFOLI gate. (b) The ripple-carry adder implemented on
  the tile lattice, with different gates depicted in different colors:
  blue tile: ID; green tile: SWAP; gold tile: TOFFOLI. Spins between
  adjacent tiles are forced to be equal by the ferromagnetic `grout'
  coupling $K$. (c) The ripple-carry adder mapped to a vertex model
  with periodic boundary condition in the transverse direction. After
  each column of gate (vertex) operation, bit states are labeled at
  each bond. Light yellow and grey stripes represent the $P$ and $T$
  matrices used in the transfer matrix calculation of the partition
  function.}
\label{fig:ripple-carry}
\end{figure}

The tiles representing the gates can then be laid down side-by-side on
a plane to implement the computational circuit, as shown in
Fig.~\ref{fig:ripple-carry} for the example of the ``ripple-carry
adder'', which computes the carry bit that is ``rippled'' to the next
bit when adding two numbers \cite{vedral}. (The ``ripple-carry adder''
is the building block for more complicated circuits such as addition
and multiplication.) As can be seen from this example, one may also
need to include the Identity (ID) gate in addition to the TOFFOLI and
SWAP gates in order to represent particular logic circuits via
tiling. Implied in the figure is that common boundaries of adjacent
tiles contain a pair of ``twin'' bits (one on each tile) whose values
must coincide. The derivations of spin Hamiltonians implementing
the truth tables of individual tiles, the short range inter-tile
Hamiltonian enforcing the consistency between bits of neighboring
tiles, and the boundary conditions specifying inputs and outputs
are presented in the Methods Section.

The final step of our mapping, also detailed in the Methods Section,
is to construct a vertex model on a tilted square lattice, with each
vertex representing either a TOFFOLI gate or four possible rectangular
tiles obtained by combining square ID and SWAP tiles (ID-ID, ID-SWAP,
SWAP-ID, SWAP-SWAP), as shown in Fig.~\ref{fig:ripple-carry}. This
construction can always be done by an appropriate retiling of the
circuit so that each rectangular tile has four neighbors (hence the
square lattice). There are six Boolean (or spin) variables associated
to each vertex: two on each of the two double bonds and one on each of
the two single bonds tied to a vertex. In deriving the vertex model we
work in the limit in which the spin coupling defining the gate
Hamiltonians, $J\to \infty$ (see the Methods Section), in which case
all gate truth tables are satisfied exactly. Consequently, each vertex
can be in one of $r=2^3=8$ states. Three of the spins are inputs, and
we use the state $q$ of the vertex, where $q=0,1,\dots,7$, to read-off
the inputs in binary (which are uniquely related to the spin): $x^{\rm
  IN}_a={\rm bit}[a,q]$, $a=1,2,3$ for the three bits of the number
$q$. The output bits are the bits of the 3-bit number $G(q)$, where
$G$ is the gate function: $x^{\rm OUT}_a={\rm bit}[a,G(q)]$,
$a=1,2,3$. The energy cost for two adjacent gates that are
incompatible with each other is determined by the ferromagnetic
coupling $K$.

The resulting vertex model Hamiltonian can be written as
\begin{eqnarray}
  \label{eq:H-vertex}
  \hat H&=& \sum_{\langle s s'\rangle} \sum_{q_s,q_{s'}}\;K^{g_s
    g_{s'}}_{q_s,q_{s'}} \;|q_sq_{s'}\rangle\langle q_sq_{s'}|
  \nonumber\\ &&+ \sum_{s\in {\rm boundary}}
  \sum_{q_s}\;h_{q_s}\;|q_s\rangle\langle q_s| \nonumber\\ &&+
  \sum_{s} \sum_{q_s,q'_{s}}\;\Delta_{q_s,q'_{s}}\;|q_s\rangle\langle
  q'_s| \;,
\end{eqnarray}
where $K^{g_s g_{s'}}_{q_s,q_{s'}}$ encodes the energy cost for
mismatched nearest-neighbor vertices (the energies, with scale set by
$K$, depend on the state of the vertices $q_s$ and $q_{s'}$, as well
as on the types of gates $g_s$ and $g_{s'}$ present at neighboring
vertices $s,s'$ -- an explicit example is given in the Supplementary
note 2); $h_{q_s}$ encodes the boundary conditions, which we
associate directly with the vertex rather than with the input or
output bits of a gate (since the relationship is one-to-one); and
finally, the transition matrix elements $\Delta_{q_s,q'_s}$ between
the states within a vertex $s$. All these couplings can be determined
given a computational circuit and the boundary conditions. The quantum
term $\Delta_{q_s,q'_{s}}$ can be designed from the internal couplings
within the tiles; For simplicity, one should consider the case,
$\Delta_{q_s,q'_{s}}=\Delta$ for all $q_s,q_{s'}$, which then
represents the 8-state counterpart of a transverse field.

The vertex model defined by Eq. (\ref{eq:H-vertex}) is the starting
point for all the subsequent discussions of this paper. For example, a
quantum annealing protocol for solving a factoring problem would start
with $K\ll \Delta$, where the ground state is a superposition of all
locally satisfied gates independent of one another, and end with $K\gg
\Delta$, with the ground state in which each tile satisfies the gate
constraint and also passes and receives the right information to and
from its neighbors.

\subsection{The quantum vertex model phase diagram}
\label{sec:annealing_results}

\begin{figure}[h!]
\centering
\includegraphics[angle=0,origin=c,width=8cm]{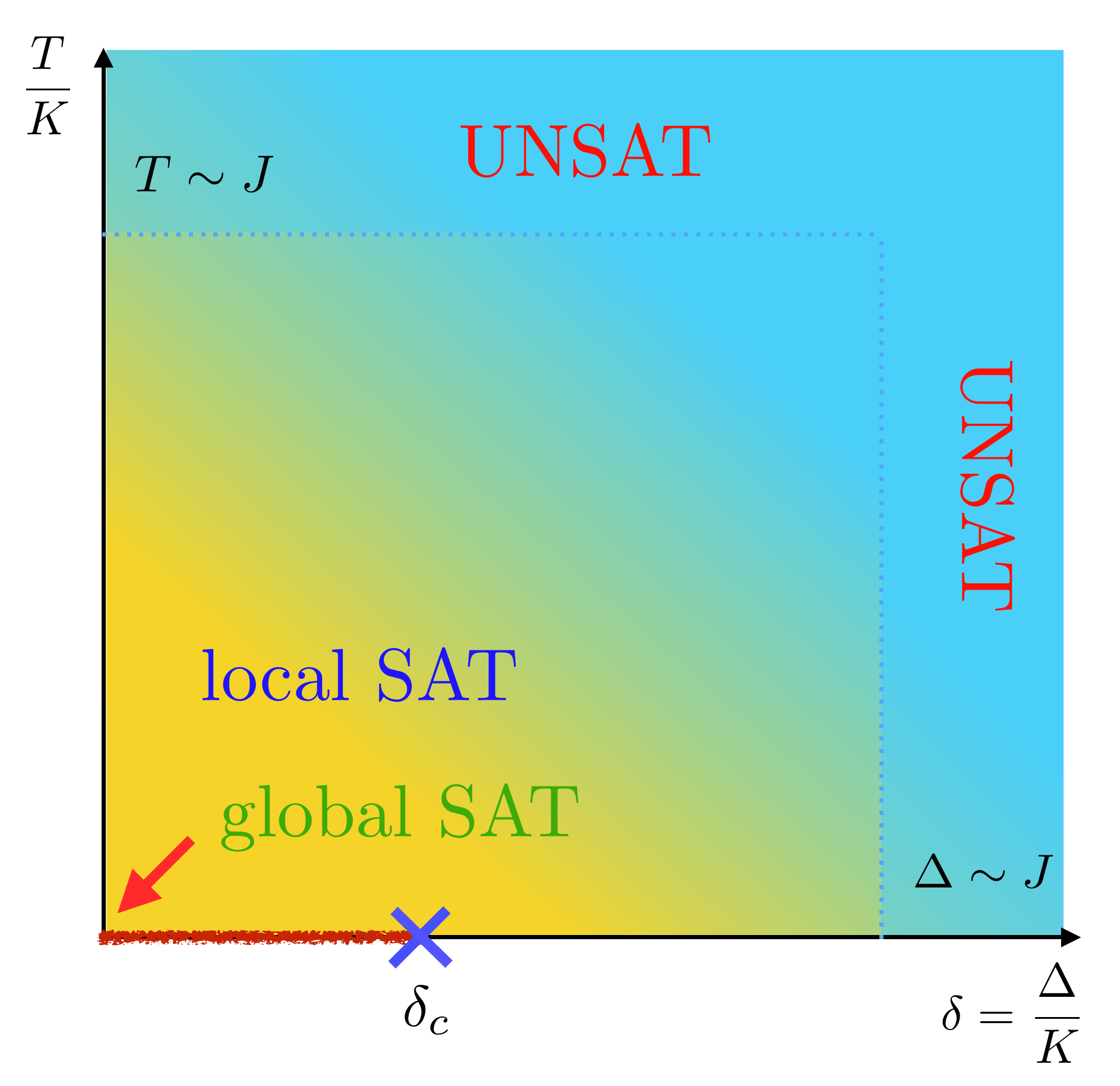}
\caption{{\bf Phase diagram of the vertex model.} Our exact calculation of the partition
  function shows that there is no phase transition along the classical
  path ($\delta = 0$). We argue that there should be a quantum phase
  transition for some critical $\delta_{\rm c}$.}
\label{fig:phase-diagram}
\end{figure}

Figure \ref{fig:phase-diagram} shows our conjectured equilibrium phase
diagram of the vertex model described by the Hamiltonian in
Eq.~(\ref{eq:H-vertex}). For $T,\Delta \ll J$, the local gate
constraints are satisfied, which we indicate by ``local SAT''. The
solution of the computational problem resides at the origin
($T/K=\Delta/K=0$), where all the gates are locally satisfied and
globally consistent, which we indicate by ``global SAT''. In the
Methods Section we show explicitly that along the classical axis,
$\delta=\Delta/K= 0$, the vertex model displays no finite temperature
bulk thermodynamic transition irrespective of the computational
circuit it represents. In particular, the resulting bulk thermodynamic
behavior is always that of a paramagnet:
\begin{equation}
  \beta F =-\left[3L(W-1)\right]\; \ln (2\cosh \beta K) \;.
\end{equation}
%

Moreover, along the ``quantum'' axis $T=0$ the vertex model must
encounter a zero-temperature quantum phase transition at a finite
value of $\delta$. This follows from considering trivial classical
circuits with no TOFFOLI gates in which case an $L\times W$ vertex
model is equivalent to $3L$ decoupled Ising chains of size $W$ in a
transverse magnetic field. Just as in the one-dimensional Ising model
in a transverse field, in the limit of no TOFFOLI gates one expects a
zero-temperature second-order quantum phase transition at $\delta_{\rm
  c} = 1$. The addition of TOFFOLI gates complicates the analysis, but
on physical grounds we expect that the phase transition cannot simply
disappear but rather change character instead, possibly from second
order to first order. This could be the case if the no-TOFFOLI
critical point happens to be an endpoint of a phase boundary in the
$\delta$-$x_{\rm T}$ plane, where $x_{\rm T}$ is the concentration of
TOFFOLI gates. Determining the order of the transition for the vertex
model describing a generic computation is a difficult problem, which
we expect to address via quantum Monte Carlo simulations in a future
publication.

\subsection{Thermal annealing of the classical vertex model}
\label{sec:numerics}

Here we study the dynamics of relaxation to the ground state as a
function of the size and depth of the computation via thermal
annealing \cite{SA}.  This proceeds by cooling the system from a high
temperature of order $K$ down to zero temperature over a total time
duration, $\tau$, according to the ramp protocol, $T(t)=K (1-t/\tau)$.

The dynamics is extracted by following an order parameter $m$
that measures the overlap of the final state $\{q^{\rm final}\}$
reached at $t=\tau$ with the reference (solution) state
$\{q^{\text{sol}}\}$:
\begin{equation}
m = \frac{8}{7} \left[\frac{1}{LW} \sum_s \delta_{q^{\rm
      final}_s,q^{\text{sol}}_s} - \frac{1}{8}\right] \;.
\label{eqn:order}
\end{equation}
(Below we explain in detail how a unique solution state
$\{q^{\text{sol}}\}$ is obtained.) Notice that the order parameter
reaches $m=1$ when the final state agrees with the solution, and $m=0$
if the state is random, in which case it agrees with the solution by
chance in 1/8th of the sites. We remark that the ``solution overlap''
is a much better indicator of the evolution towards solution than the
total energy. This is because a single vertex flip into an incorrect
state in the middle of the circuit may cost little energy but it
throws other vertices into a completely different state from the
correct one.

\begin{figure*}[ht]
\centering
\includegraphics[angle=0,origin=c,width=18cm]{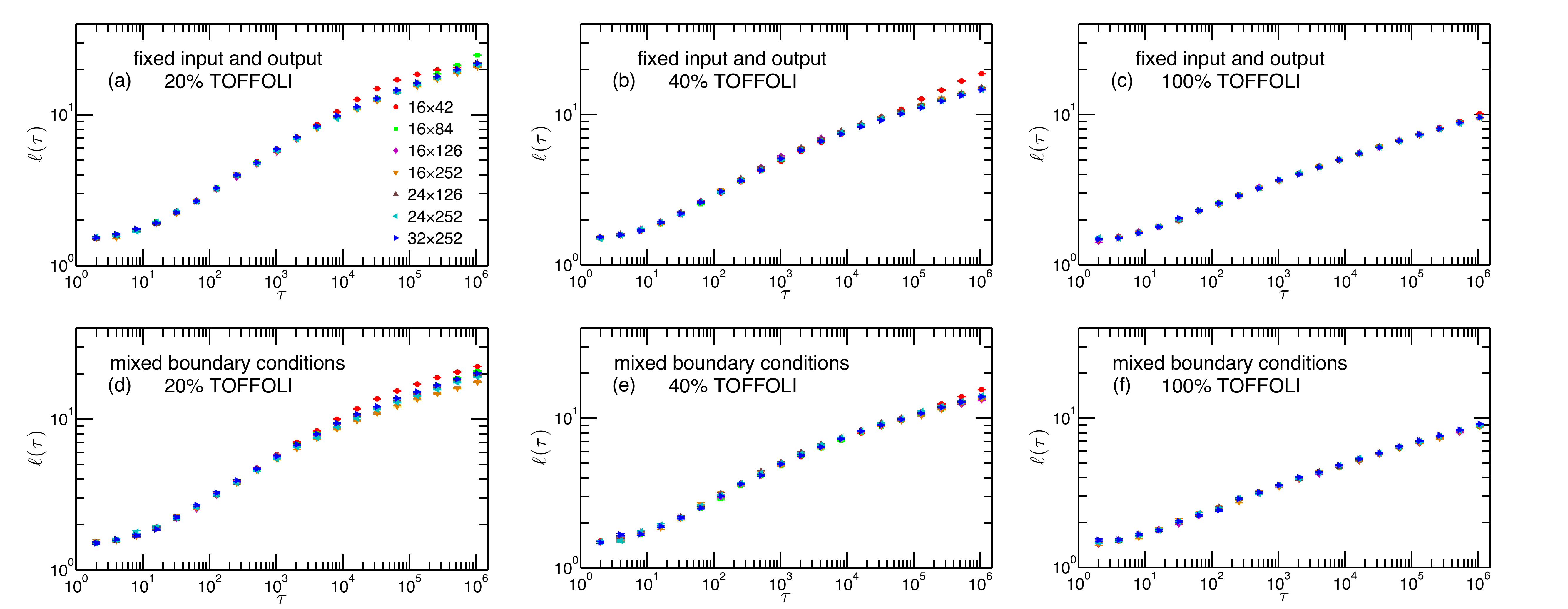}
\caption{{\bf Scaling of the dynamical correlation length $\ell(\tau)$}. For all sizes and cases, 2000 realizations of the
  boundary states were used. The data-point code used in panel (a)
  applies to all other panels. When not visible, the error bars are
  smaller than the size of the data points. (a)-(c) fixed input and
  output; (d)-(f) mixed boundary conditions; (a)\&(d) 20\% TOFFOLI;
  (b)\&(e) 40\% TOFFOLI; (c)\&(f) 100\% TOFFOLI. For systems with the
  smallest depth $W$ studied, $16\times 42$, and for circuits with few
  TOFFOLI gates (20\%) and fixed input and output boundary conditions,
  $\ell(\tau)$ tends to saturate, indicating that complete solutions
  have been reached. Notice that the functional form of the scaling
  does not depend on the boundary conditions, and depends solely on
  the concentration of TOFFOLI gates.}
\label{fig:MC_scaling_plot1}
\end{figure*}

\begin{figure}[h!]
\centering
\includegraphics[angle=0,origin=c,width=8.5cm]{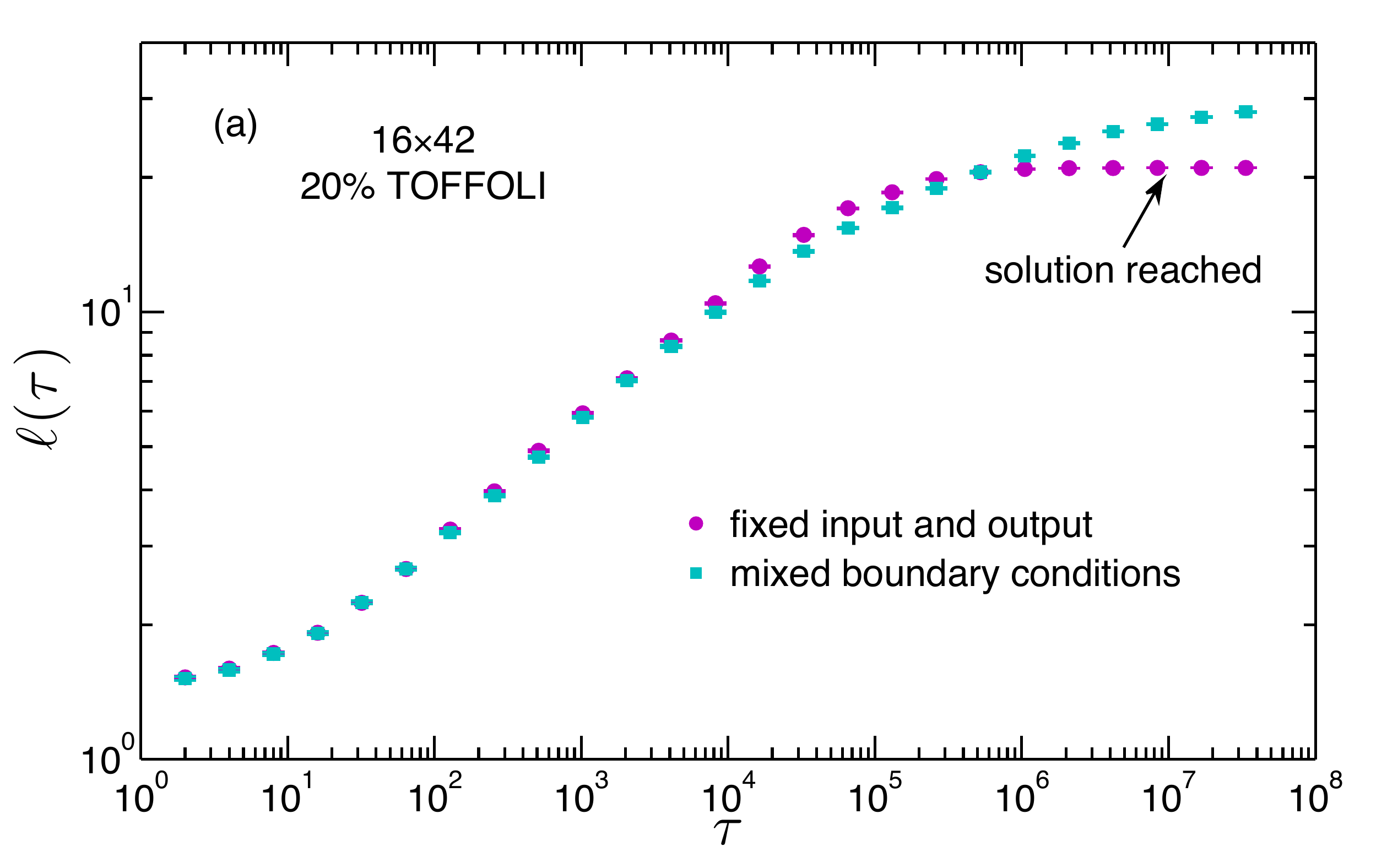}
\includegraphics[angle=0,origin=c,width=8.5cm]{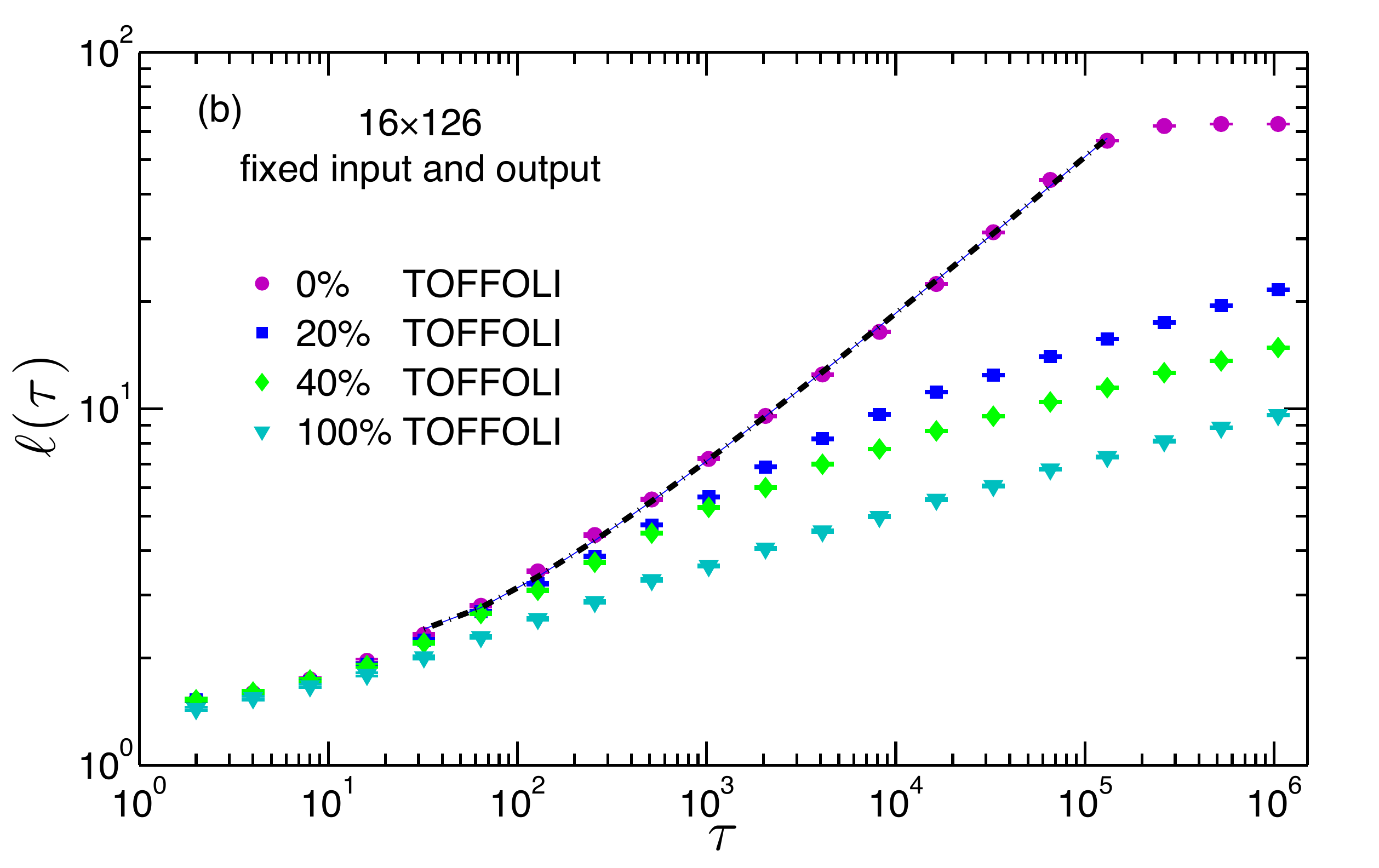}
\caption{{\bf Effects of boundary conditions and TOFFOLI concentration.} (a) The saturation of the scaling curve for fixed input and
  output boundary conditions with 20\% TOFFOLI gates when
  $\ell(\tau)\sim W/2$. This indicates that the solutions have been
  reached, and is consistent with the domain growth picture. Notice
  that for mixed input and output boundary conditions, the curve does
  not saturate when $\ell(\tau)\sim W/2$. (b) The functional form of
  the scaling curves as a function of different TOFFOLI
  concentrations. The correlation grows more slowly as the
  concentration of TOFFOLI gates increases. The dashed line
  corresponds to the fitting $\ell(\tau)=\ell_0[(\tau/\tau_0)/{\rm
      ln}(\tau/\tau_0)]^{1/2}$, with $\ell_0=1.42$ and $\tau_0=8.33$.}
\label{fig:saturation}
\end{figure}

The details of the numerical Metropolis simulations are presented in
the Methods Section. Our results are represented in the form inspired
by the dynamic scaling theory of Ref.~\cite{Liu-Polkovnikov-Sandvik}
that builds on the Kibble-Zurek mechanism~\cite{kibble,zurek}, namely:
\begin{equation}
\ell(\tau) = \langle m \rangle (\tau)\; WL/L_{\partial} \;,
\label{eqn:scale}
\end{equation}
which defines a dynamical correlation length,
$\ell(\tau)$. $L_{\partial}$ is the number of pinned vertices on both
boundaries (see the Methods Section). To motivate
Eq.~(\ref{eqn:scale}) we note that the domain of satisfied gates that
contribute to $\langle m \rangle (\tau)$, the fraction of gates that
reach their correct states at time $\tau$, grows from the pinned
states at the boundaries, and covers an area $L_{\partial}\times
\ell(\tau)$. Thus $\ell(\tau)$ describes the growth of correlated
regions of satisfied gates that eventually connect the two boundaries
of the circuit. (We note that recently, the Kibble-Zurek mechanism has
been extended to include systems with zero-temperature
order~\cite{Rubin}, the case relevant to the current discussion).

We note that at any temperature $T$ along the annealing path, the
correlation length is $\ell_T (\tau) \leq \ell _T (\tau \to \infty) =
\xi_T$, where $\xi_T\sim e^{K/2T}$ is the thermal correlation length
in the paramagnetic state, and $K$ is a characteristic ferromagnetic
interaction strength in our model. In thermal equilibrium all gate
constraints defining the computational circuit would be satisfied once
$\xi_T$ reaches the depth of the computation, $W$.  Notice that the
exponential dependence of $\xi_T$ on temperature implies that
achieving the correct assignment of gates does not require very low
temperatures on the scale of $K$ since $\xi_T \sim W$ already for
temperatures below $T\sim K/\ln W$. However, reaching the solution to
the computational problem is a dynamical process that cannot proceed
to completion until the {\it dynamic correlation length} at the end of
the annealing protocol, $\ell (\tau) = \ell_{T=0} (\tau)$, reaches
$W$, allowing the input and output boundaries of the system that
specify the computation to communicate.

In Fig.~\ref{fig:MC_scaling_plot1} we present the numerical results
for fixed input and output, and mixed boundary conditions, with
different concentrations of TOFFOLI gates (see the Methods Section for
details). Remarkably, we find that the curves for different system
sizes $L$ and $W$ collapse very well when scaled as in
Eq.~(\ref{eqn:scale}). In addition, notice that for shorter circuit
with fixed input and output and low concentration of TOFFOLI gates
(20\%), $\ell(\tau)$ begins to saturate for large enough $\tau$
(Fig.~\ref{fig:MC_scaling_plot1}a). As shown more clearly in
Fig.~\ref{fig:saturation}a, this saturation occurs when the dynamical
correlation length $\ell(\tau)$ reaches $W/2$, where the growing
domains of satisfied gates meet. Since in this case $L_{\partial} =
2L$, $\ell(\tau_s) \sim W/2$ corresponds to $\langle m \rangle
(\tau_s) = 1$ establishing $\tau_s$ as the time-to-solution.  For
mixed boundary conditions, however, $\ell(\tau)\sim W/2$ initiates the
communication between the two boundaries and establishes the system's
capacity to ``learn'' (see below) but is not sufficient for
negotiating solution. Indeed, Fig.~\ref{fig:saturation}a shows that
$\langle m \rangle (\tau)$ does not yet saturate when $\ell(\tau)\sim
W/2$.  As can be seen from Fig.~\ref{fig:correlation_growth} for
computations with mixed boundary conditions, correlations must develop
along the transverse direction ({\it i.e.}, parallel to the
boundaries) before solution can be reached. In those cases it is this
slower process that determines the time-to-solution and dominates the
complexity of computations.

Finally, all non-trivial operations between input and output bits
involve TOFFOLI gates, and it is thus expected that the increasing the
concentration of these gates slows down the growth of
correlations. This expectation is confirmed in
Fig.~\ref{fig:saturation}b, where we show curves for the same system
size with different concentrations of TOFFOLI gates. The case of no
TOFFOLI gates is equivalent to $3L$ decoupled ferromagnetic Ising
chains. In this case the dynamic correlation length behaves as
$\ell(\tau)=\ell_0[(\tau/\tau_0)/{\rm ln}(\tau/\tau_0)]^{1/2}$ (with
$\ell_0=1.42$ and $\tau_0=8.33$) as illustrated by the dashed line in
Fig.~\ref{fig:saturation}b. This behavior is in agreement with the
exact result for the Kibble-Zurek dynamical scaling of the density of
domain walls in a ferromagnetic Ising chain \cite{Krapivsky}.

\subsection{Annealing with learning}

\label{sec:learning}

\begin{figure*}[hbt]
\centering
\includegraphics[angle=0,origin=c,width=18cm]{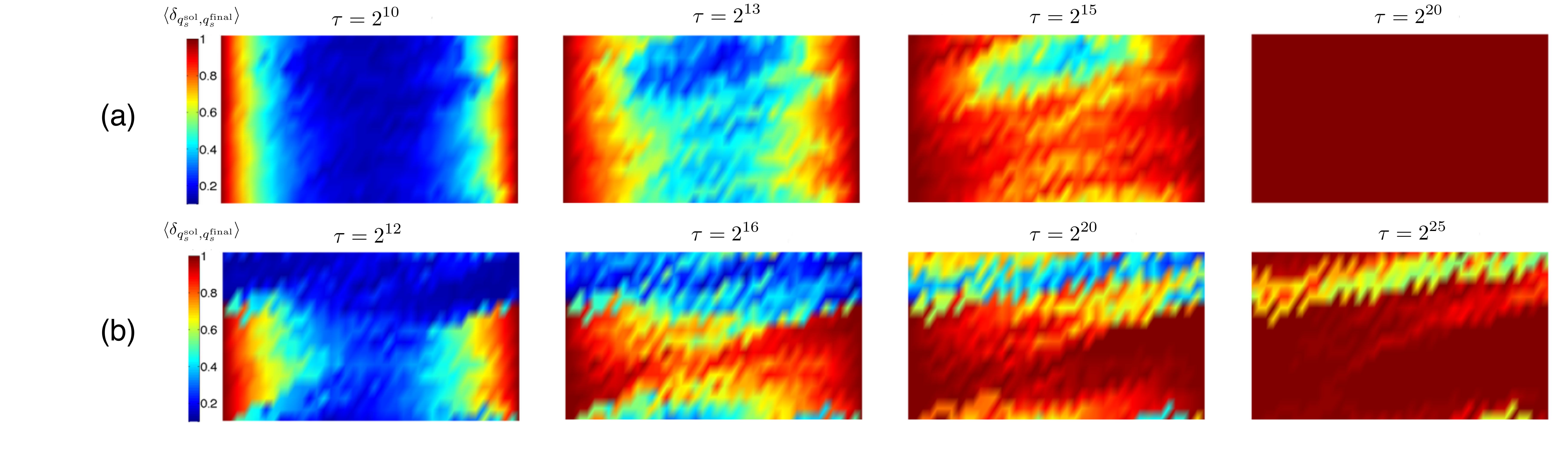}
\caption{{\bf Growing correlation length without learning.} The average local overlaps $\langle\delta_{q^{\rm sol}_s,
    q^{\rm final}_s}\rangle$ of 2000 replicas with $\{q^{\rm sol}\}$
  for a given circuit and boundary state without learning. The system
  size is $16\times 42$, with 20\% TOFFOLI gates. (a) Fixed
  input and output; (b) mixed boundary conditions.}
\label{fig:correlation_growth}
\end{figure*}

Simple thermal annealing is not necessarily an optimal way to reach
the ground state. For example, in the case of forward computation, the
time scale for the dynamic correlation length to grow to $\ell(\tau)
\sim W/2$ (so as to reach solution) is slower than ballistic (or
linear in $\tau$), as expected for deterministic forward
computation. This can be already seen from the exactly solvable case
with no TOFFOLI gates. Moreover, for single-solution problems with
mixed boundary conditions the growth of correlations establishing
communication between boundaries scales with the same form as in
direct computation (see Fig.~\ref{fig:MC_scaling_plot1}). However, in
that case negotiating solution requires the establishment of much
slower correlations {\it along} the boundaries, a process for which a
``vanilla'' thermal annealing approach is extremely inefficient and
would require unreasonably large computational resources.

These shortcomings are addressed by using a heuristic ``learning''
protocol in which annealing proceeds
through the following steps: (1) one starts by annealing $N_R$
identical replicas of a circuit over some time $\tau_a$, during which
the correlation lengths grow beyond a few columns of gates such that
the probability for assigning correct gates within that region, $p\sim
\exp -|x|/\xi > 1/2$, within each replica; (2) one then assigns a
specific identity to each gate (with $p >1/2$) provided that a
fraction of the $N_R$ replicas, greater than or equal to $\alpha$,
agree on this assignment; (3) with the agreed upon gates frozen, the
annealing process is independently applied again to each of the
replicas allowing only gates not yet fixed to participate in the
Metropolis algorithm; finally, (4) the procedure is iterated until all
gates are fixed, thus establishing the solution to the problem.

This protocol raises the question of how many replicas $N_{R\epsilon}$
are needed to ensure that the learning algorithm reaches the correct
result with a probability greater than $1-\epsilon$. In particular,
how does $N_{R\epsilon}$ depend on the system size $L\times W$ and the
threshold $\alpha$? As we show in the Supplementary note 3, the
number of replicas needed to ensure an error rate smaller than
$\epsilon$ is given by $N_{R\epsilon} = \frac{{\rm ln} \left[
    \frac{2p-1}{p} \frac{\epsilon}{LW} \right]} {{\rm ln} \left[
    2p^{1-\alpha}(1-p)^{\alpha}\right]}$, where $p>\frac{1}{2}$ is the
probability of a correct gate assignment for one replica. Note that,
for fixed $\alpha$ and error rate $\epsilon$, the number of replicas
grows only \textit{logarithmically} with the system size, and thus in
practice the learning algorithm works with reasonable resources.

Before describing the results of applying ``annealing with learning''
to computations with both fixed and mixed boundary conditions, 
in Fig.~\ref{fig:correlation_growth} we plot the average
local overlaps of 2000 replicas with the solution for a fixed circuit
and boundary condition before applying the learning algorithm. The
agreement with the data presented in Fig.~\ref{fig:MC_scaling_plot1}
substantiates the fact that the local majority rule implemented
through the independent annealing of the replicas recapitulate the
behavior of the correct solution to the computational problem.

\begin{figure*}[ht]
\centering
\includegraphics[angle=0,origin=c,width=18.5cm]{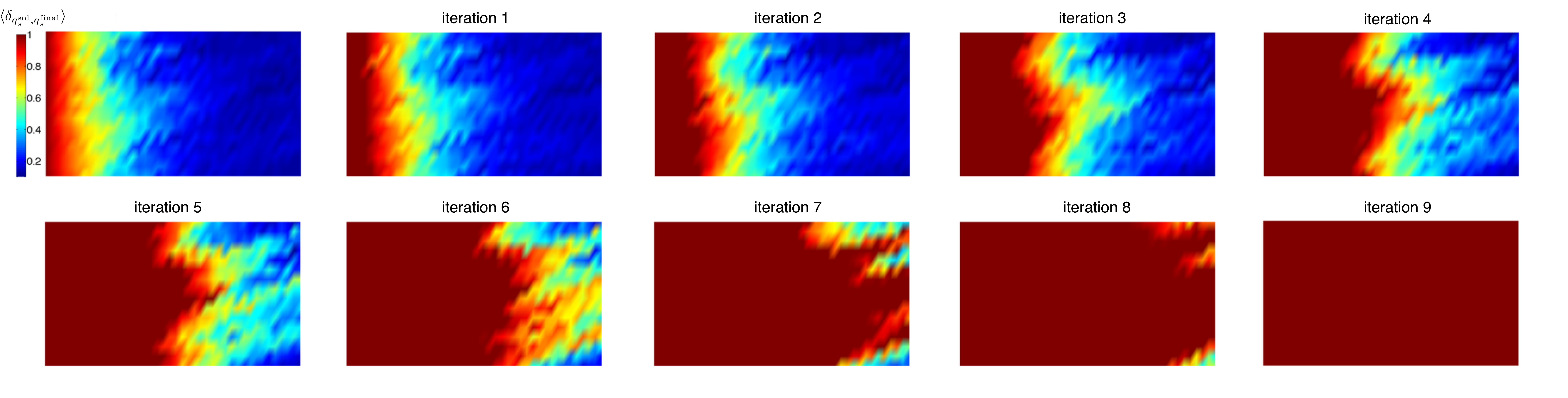}
\caption{{\bf Growing correlation length with learning for fixed input boundary condition.} Annealing with learning for the fixed input boundary case for
  a system of size $16\times 42$ with 20\% TOFFOLI gates. The
  annealing time within each iteration is $\tau_a=2^{13}$ and the gate
  state probability threshold $\alpha=0.7$.}
\label{fig:learning_forward}
\end{figure*}

We start from fixed input boundary condition. Using the algorithm described above,
we choose $\tau_a=2^{13}$ for each iteration, and set the majority
rule threshold at $\alpha=0.7$. In Fig.~\ref{fig:learning_forward} we
show the local order parameter of the final states averaged over 2000
replicas after each iteration. We emphasize that even though we are
plotting the average overlap with the actual solution as a benchmark,
in the learning algorithm no reference to $\{q^{\rm sol}\}$ is made. The
weight of each possible state of each gate in the circuit is computed
solely from the replicas. After each iteration, with $\tau_a=2^{13}$
the correlation length grows to $\ell(\tau_a)\sim 10$, and by pinning
gates with high percentage of agreement on certain states we are
pushing the ``boundary'' forward until all gates are fixed. Since the
total number of iterations $n_a$ scales linearly with the circuit
depth $W$, $n_a \propto W$, the total time to solution
$\tau=n_a\tau_a$ also scales linearly with $W$, $\tau \propto W
\tau_a$, consistent with the expectations for the time-to-solution for
forward computation. For the computation shown in
Fig.~\ref{fig:learning_forward}, it is clear that the ``annealing with
learning'' process proceeds ballistically and reaches solution with
$n_a=9$ steps.

\begin{figure*}[ht]
\centering
\includegraphics[angle=0,origin=c,width=18cm]{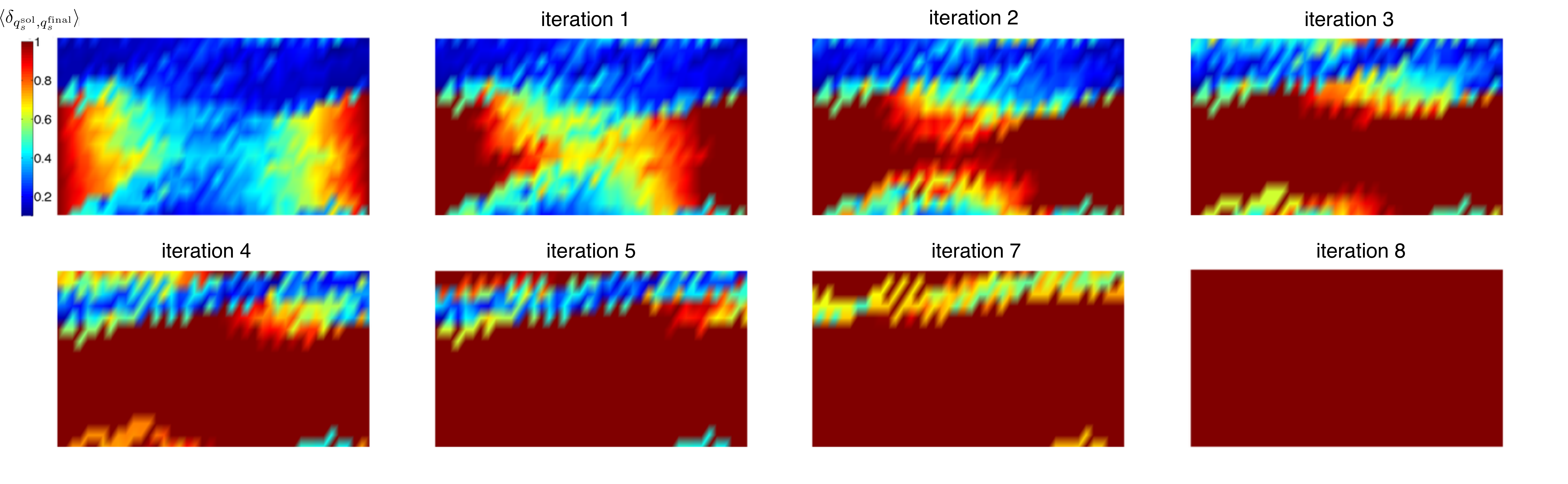}
\caption{{\bf Growing correlation length with learning for mixed boundary conditions.} Annealing with learning for mixed boundary conditions and
  systems of size $16\times 42$ with 20\% TOFFOLI gates. The annealing
  time within each iteration is $\tau_a=2^{13}$, and the probability
  threshold $\alpha=0.7$. After iteration 6 (not shown), the correlations fully build up along the
  longitudinal direction, $\tau_a$ is then increased to $\tau_b=2^{18}$.}
\label{fig:learning_mixed}
\end{figure*}

Now we look at mixed boundary conditions. The results presented in
Fig.~\ref{fig:learning_mixed} are obtained by applying the learning
algorithm with $\alpha=0.7$. However, in the case of mixed boundary
conditions the process of ``learning'' proceeds through two series of
annealing steps with different time scales: an initial set of
iterations with $\tau_a = 2^{13}$ which build longitudinal
correlations required for learning, followed by a set of longer
annealing steps with $\tau_b = 2^{18}$ that allow the slower
correlations along the transverse direction to
develop. Figure~\ref{fig:learning_mixed} shows the progression to
solution, which could not be reached for the same computation using
the ``vanilla'' thermal annealing for our longest accessible times
($\tau \sim 2^{25}$).

We note that this protocol can also be used to solve problems with a
``few'', ${\cal O}$(1), solutions. This is best illustrated for the
case of two solutions, which can be addressed by carrying out $2n$
computations with mixed boundary conditions, where $n$ is the number
of unknown bits in the input. The idea is to define $2n$ problems by
fixing each bit at a time to be 0 or 1, while leaving the other $n-1$
bits floating. Since the two solutions must differ in at least one of
the $n$ bits, after at most $2n$ steps, this scheme transforms the
problem into two separate problems, each of which can be solved by the
techniques discussed in this paper. An important problem that falls
precisely within this case is factorization of semi-prime numbers
$s=p\times q$, where there are exactly two solutions, corresponding to
the two ordered pairs $(p,q)$ and $(q,p)$ of primes $p,q$ (assumed to
be different).

\begin{figure}[ht]
\centering
\includegraphics[angle=0,origin=c,width=9cm]{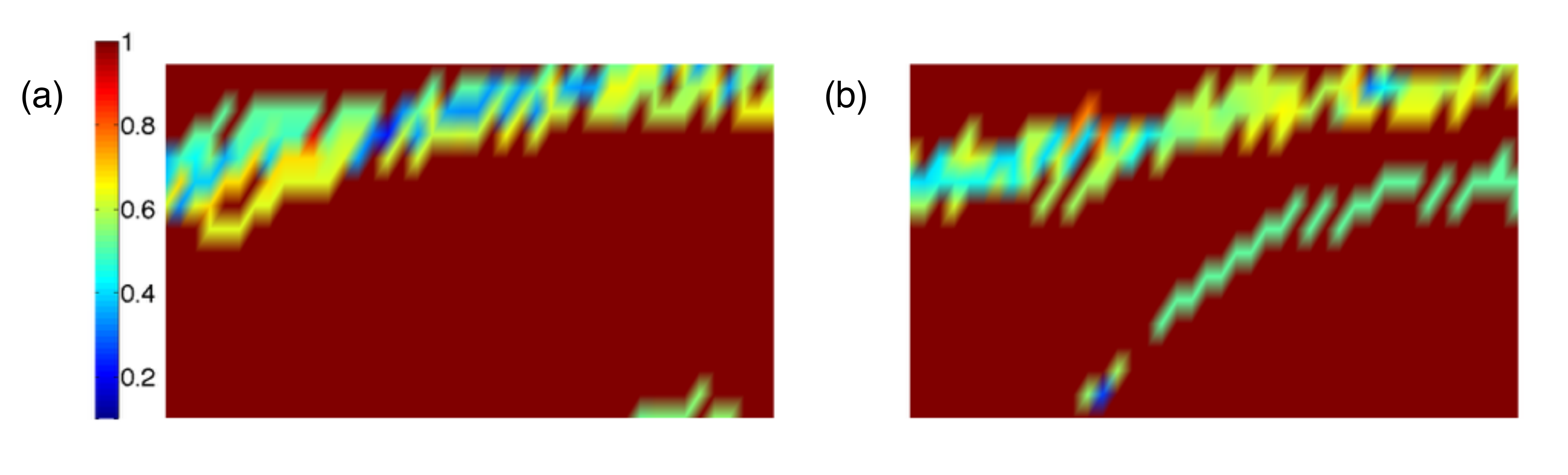}
\caption{{\bf Cases with multiple solutions or no solution.} Color plot of the largest weight of each gate state in the
  circuit for a system of size $16\times42$ and 20\% TOFFFOLI gates
  after a relaxation time $\tau\approx 2^{20}$. (a) Case with
  8 solutions; (b) case with no solution. The learning
  algorithm eventually gets stuck at the point where no more gates
  have majority weight above the threshold $\alpha$.}
\label{fig:multi_no_sol}
\end{figure}

Finally, we turn to the analysis of cases with multiple solutions and
no solution. In both of these cases it is not sensible to compute the
local overlap with a solution, as we did for circuit problems with
only one solution. Instead, we plot the largest weight of each gate
state in the circuit obtained from 2000 replicas. This is shown in
Fig.~\ref{fig:multi_no_sol} for an instance with 8 solutions obtained
by fixing fewer gates (than in the single solution case) on each
boundary; and an instance with no solutions, obtained by fixing a few
gates on one boundary to the wrong states. Fig.~\ref{fig:multi_no_sol}
shows that the learning algorithm eventually gets stuck when the
replicas cease to agree on gate assignments above the threshold
$\alpha$. We note that the learning algorithm cannot differentiate
between these two cases. We interpret the freezing of the system as an
effect of frustration in satisfying the local gate constraints in the
bulk induced by incompatible boundaries in the case of no solution or
compatible but competing boundaries in the case of multiple solutions.

\subsection{Mapping onto the D-Wave Chimera graph for quantum annealing}
\label{sec:chimera}

\begin{figure*}[!ht]
\centering
\includegraphics[angle=0,origin=c,width=0.8\textwidth]{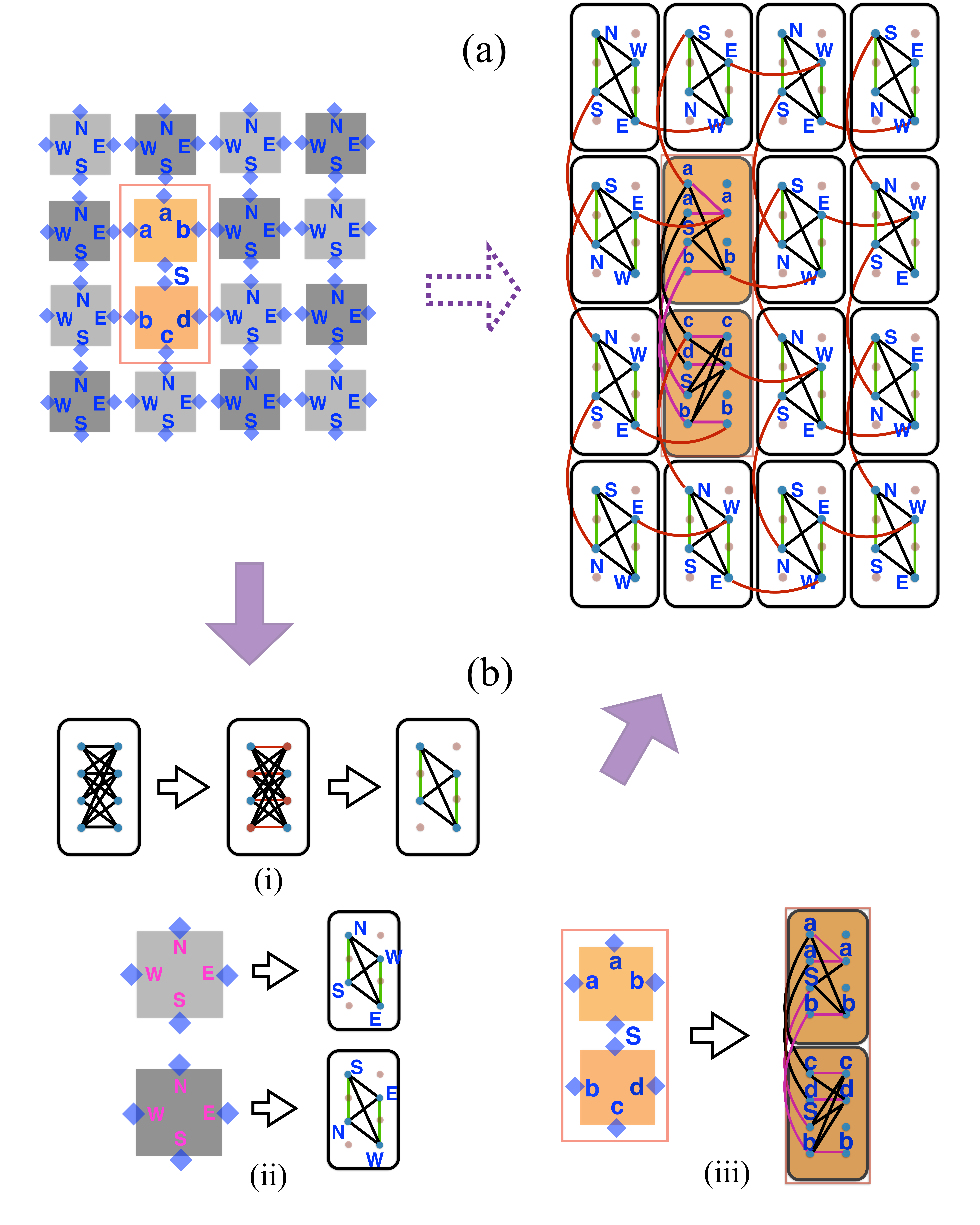}
\caption{{\bf Mapping onto the Chimera graph.} Procedure for embedding a $4\times 4$ tile lattice into the
  Chimera graph. (a) Left: a generic tile lattice rotated by
  $45^\circ$. Spins are put on the boundary of each tile. The lattice
  can be further divided into two sublattices, depicted by dark and
  light grey respectively; right: embedding of the tile lattice into
  the Chimera graph. The ``grout couplings'' are indicated by red
  links. (b) Embedding of each gate into the unit cells of the Chimera
  graph. (i) Left: a $K_{4,4}$ unit cell of the Chimera graph; middle:
  in order to couple qubits in the same column, we slave the qubits to
  their neighbors in the other column using additional ferromagnetic
  couplings indicated by red links; right: effectively we are left
  with four qubits that are fully connected. For simplicity, we
  hereafter denote the effective couplings between spins in the same
  column by a single green link. However, one should keep in mind that
  they are obtained by slaving the spins to the opposite column via
  large ferromagnetic couplings. (ii) The four qubits in the rotated
  square tile are labeled by their locations on the tile: N (North), S
  (South), W (West) and E (East). Tiles corresponding to different
  sublattices must be embedded differently due to the special
  connectivity of the Chimera graph. (iii) Embedding of the TOFFOLI
  gate consisting of two square tiles into two unit cells. $(a, b, c,
  d)$ corresponds to the input and output bits of the gate, and $S$ is
  the ancilla bit. In the unit cell, ferromagnetic couplings that copy
  spins are indicated by purple links, and couplings required in
  Hamiltonian (\ref{H-Toffoli}) are indicate by black links.}
\label{fig:chimera}
\end{figure*}

We close this paper by describing a scheme for ``programming'' our
vertex model into a quantum annealer. In particular, we present an
explicit embedding of the tile model of universal classical computing
circuits into the Chimera graph architecture of the D-Wave
machine. The idea is to use one unit cell to represent one square tile
of our construction presented previously. Rectangular tiles
(\textit{i.e.}, TOFFOLI gates) can be viewed as consisting of two
square tiles, thus requiring two unit cells to be embedded in the
Chimera graph. We then implement the Hamiltonians of
Eqs. (\ref{H-ID})-(\ref{H-Toffoli}) using the programmable couplers
available in the D-Wave machine, as illustrated in
Fig.~\ref{fig:chimera} and described in more detail in the Methods
Section.

\section{Discussion}
\label{sec:conclusions}

The results of this paper were motivated by an attempt to use our
statistical mechanics intuition about lattice models of spin systems
to uncover some of the salient features of universal classical
reversible computation. There are questions posed and open problems
raised by these studies. Here we list four that we find most
important.

First, one should understand the scaling of time-to-solution of the
various schemes discussed here, including those that utilize learning,
as a function of input size and depth for specific computational
problems. Under a trivial reduction scheme, one can solve problems
with two solutions using similar annealing with learning techniques
that we deployed for problems with a unique solution. As an important
application we are already investigating the problem of the
factorization of semi-primes. The scaling properties of the
time-to-solution in the context of this concrete and relevant problem
should be contrasted to that obtained in the random circuit with the
same concentration of TOFFOLI gates.

A second question raised by our work is the nature of the
zero-temperature quantum phase transition encountered in the quantum
vertex model, as depicted in Fig.~\ref{fig:phase-diagram}. We
demonstrated that, in the limit of the trivial computational circuit
with no TOFFOLI gates, this transition is second order, in direct
analogy to the case of the one-dimensional Ising model in a transverse
magnetic field. Whether the transition remains second order or becomes
first order for realistic computations (corresponding to a finite
concentration of TOFFOLIs) has very important consequences for
solutions of computational problems via quantum annealing.

Third, the computational problems discussed here should also be
studied directly in a {\it bona fide} quantum annealer. An important
result of this paper is the programming of generic reversible
computational circuits into the Chimera architecture of the D-Wave
machine. This paves the way for using this type of hardware to study
annealing protocols along the $\delta$ axis, as well as arbitrary
directions in the $\delta - T$ plane. Our approach should also be used
as a guide to the development of alternative machine architectures
optimized for direct implementations of the vertex model.

Finally, we close with a brief discussion of the broader implications
of the mapping of reversible classical computation onto the vertex
model on the individual disciplines of computer science and
physics. As already mentioned earlier, the line of argumentation in
this paper follows a physics perspective, namely, it concentrates on
``typical behavior'' based on heuristic approach to explicit
instantiations of the vertex model. Computer science could benefit
from further work on more sophisticated theoretical and computational
heuristic approaches, special purpose hardware ({\it i.e.}, quantum
annealers), and new formal proofs that rely on statistical mechanics
representations of computational problems. At the same time there are
lessons to be learned from computer science that we believe may have
interesting implications for physics. For example, if NP$\neq$P, the
vertex model representing the hardest problems in UNIQUE-SAT can be
also viewed as describing a physical glassy system that displays slow
dynamics even though the model involves no frustrating interactions,
has a unique non-degenerate ground state, and displays no bulk
thermodynamic transitions down to zero temperature! There are known
examples of systems with glassy dynamics in the absence of a
thermodynamic phase transition, such as the kinetically constrained
models discussed in
\cite{Newman-Moore1999,Garrahan-Newman2000,Ritort-Solich2003}.
However, the non-Arrhenius relaxation characteristic of these models
only translate into a quasi-polynomial time-to-solution of a
computational problem. Thus, within the vertex model approach, the
existence of hard UNIQUE-SAT problems with exponential or
sub-exponential behavior of the time-to-solution would suggest the
existence of a novel family of glassy physical systems without a
thermodynamic transition but with exponentially large barriers and
corresponding astronomically-long relaxation times.  This example
underscores the richness of the possibilities opened by explorations
of the vertex model of classical computation and more generally, of
problems at the interface between physics and computer science.


\section{Methods}

\subsection{Implementing gates with one- and two-body spin interactions}
\label{sec:model-definition}

We start by representing Boolean variables $x_i= (1+\sigma_i)/2$ in
terms of spins $\sigma_i=\pm 1$ placed on the boundary of each tile,
as depicted in Fig~\ref{fig:gate_tile_coupling}. Operations of logic
gates are then implemented in a similar way as in
Ref.~\cite{Biamonte2008}, by designing a Hamiltonian acting on the
spins associated with individual tiles such that (a) the interactions
are short ranged and involve at most two bodies; and (b) spin (i.e.,
bit) states that satisfy the gate constraint are ground states of the
tile Hamiltonian and all other ``unsatisfying'' spin-states are pushed
to high energies.

\noindent
{\textbf{Identity (ID) Gate:}} The ID gate takes two bits $(a,b)$ into
$(a,b)$. This is easily enforced by adding ferromagnetic interactions
($J>0$) that align input bits $a$ and $b$ to output bits $c$ and $d$,
respectively, leading to an energy
\begin{equation}
E_{{\rm ID}}(\sigma_a, \sigma_b; \sigma_c, \sigma_d) = - J(\sigma_a
\sigma_c+\sigma_b \sigma_d).
\label{H-ID}
\end{equation}

\noindent
{\textbf{SWAP Gate:}} The SWAP gate takes $(a,b)$ into $(b,a)$, and
can be implemented in the same manner as the ID gate through a
ferromagnetic interaction ($J>0$),
\begin{equation}
E_{{\rm SWAP}}(\sigma_a, \sigma_b; \sigma_c, \sigma_d) = - J(\sigma_a
\sigma_d +\sigma_b\sigma_c).
\label{H-SWAP}
\end{equation}

\noindent
{\textbf{TOFFOLI Gate:}} The TOFFOLI gate is represented by a
rectangular tile with the three input bits $(a, b, c)$ and three
output bits $({a'},{b'},d)$ placed on the boundary, as shown in
Fig.~\ref{fig:gate_tile_coupling}. Notice that in this case we also
place an additional ancilla bit in the center of the rectangular tile,
which is essential in order to satisfy the gate constraint with no
more than two-body interactions. The TOFFOLI gate takes the three-bit
input state $(a,b,c)$ into $(a,b,ab\oplus c)$. The copying of the
first two input bits from the input into the output is accomplished as
before through a ferromagnetic coupling: $-J(\sigma_a
\sigma_{a'}+\sigma_b \sigma_{b'})$. Enforcing the third output bit
$d=ab\oplus c$ requires a more involved interaction. We present the
result below, and leave the detailed justification for the
Supplementary note 1. The complete energy cost associated to the
TOFFOLI gate reads
\begin{widetext}
\begin{eqnarray}
E_{{\rm TOFFOLI}}(\sigma_a, \sigma_b, \sigma_c; \sigma_{a'},
\sigma_{b'}, \sigma_{d}; \sigma_S) & = & -
J(\sigma_a\sigma_{a'}+\sigma_b\sigma_{b'}) + J(\sigma_a -3\sigma_b -
2\sigma_c +2\sigma_d +4\sigma_S) \nonumber \\ & &
+\ J(-3\sigma_a\sigma_b - 2\sigma_a\sigma_c + 4\sigma_b\sigma_c +
2\sigma_a\sigma_d-4\sigma_b\sigma_d - 4 \sigma_c\sigma_d \nonumber
\\ & & \;\;\;\;\;\;
+\ 4\sigma_a\sigma_S - 8\sigma_b\sigma_S - 6\sigma_c\sigma_S +
6\sigma_d\sigma_S).
\label{H-Toffoli}
\end{eqnarray}
\end{widetext}
%

\subsection{The global constraint and coupling of adjacent tiles}
\label{sec:tile-coupling}

In addition to satisfying each gate separately, spins shared by
neighboring tiles must be matched across the entire system in order
for the tile model to accurately represent the desired computational
circuit. To be precise one can imagine splitting each boundary spin
into two ``twin'' spins and identifying input/output spins with each
tile. Within this picture, adjacent spins at the boundary between
tiles must be locked together, a constraint we implement by
introducing a ferromagnetic ``grout'' coupling $K>0$ between spins on
adjacent tiles. The corresponding term in the energy is then written
as
\begin{equation}
E_{\rm grout}(\{\sigma\}) = - K \sum_{\langle i,j\rangle}
\sigma_i\;\sigma_j,
\end{equation}
where $\langle i,j\rangle$ labels pairs of ``twin'' spins $i$ and $j$
on the boundary between two adjacent tiles and the sum ranges over all
such pairs of the system.

\subsection{Boundary conditions}
\label{sec:boundary_conditions}

Completing the description of the two-dimensional model of universal
classical computation requires a discussion of boundary conditions,
which determine the type of computational problem one is addressing.
For example, if the $N$-bit input is fully specified and one is
interested in the output, all that is needed is to transfer the
information encoded into the input left to right by applying
sequentially the gates one column of tiles at a time. In this case, if
the depth (i.e., the number of steps) of the computation is a
polynomial in $N$, this column-column computation reaches the output
boundary, and thus solves the problem, in polynomial time.

As mentioned earlier, by using reversible gates one can also represent
computational problems with mixed input-output boundary conditions for
which only a fraction of the bits on the left (input) edge and a
fraction of the bits on the right (output) edge are fixed. A concrete
example is the integer factorization problem implemented in terms of a
reversible integer multiplication circuit. A reversible circuit for
multiplying two $N$-bit numbers $p$ and $q$ can be constructed using
$5N+1$ bits in each column. One needs two $N$-bit registers for the
two numbers $p$ and $q$ to be multiplied, one $N$-bit carry register
$c$ for the ripple-sums, a $2N$-bit register $s$ for storing the
answer $p\times q=s$, and one ancilla bit $b$. For multiplication, one
only fixes the boundary conditions on the input: $p$ and $q$ are the
two numbers to be multiplied, and $c$, $s$ and $b$ are all 0's. For
factorization we must impose mixed boundary conditions: On the input
side the $c$, $s$ and $b$ registers are fixed to be all 0's; on the
output side the $s$ register is now fixed to the number to be
factorized, and $c$ and $b$ are again all set to 0. Thus, $3N+1$ bits
in the input and output are fixed, while $2N$ bits are floating on
both boundaries.

Boundary conditions on inputs, outputs, or both are imposed by
inserting longitudinal fields at the appropriate bit sites, namely,
\begin{equation}
E_{\rm boundary}(\{\sigma\}) = -\!\!\!\sum_{i\in{\rm boundary}} h_i\,
\sigma_i,
\end{equation}
with $|h_i| = h \gg J$. The sign of an individual $h_i$ field
determines the value of the spin $\sigma_i$ and thus of the binary
variable $x_i$: For $h_i>0$, $x_i=1$, while for $h_i<0$, $x_i=0$. If
no constraint is imposed on a binary variable $x_i$, then $h_i=0$.

\subsection{Construction of the vertex model}
\label{sec:strong_coupling} 
        
Combining the contributions above leads us to a classical Hamiltonian
that includes the energy functions internal to each tile, the coupling
between the spins at the boundary between adjacent tiles, and the
magnetic fields associated with the input and output bits defining the
boundary conditions of the computation, namely,
\begin{equation}
\label{eq:E_C}
H_C = \sum_{g} E^J_g (\{\sigma\}_g) - K \sum_{\langle i,j\rangle}
\sigma_i\, \sigma_j
- \!\!\!\sum_{i\in \text{boundary}} h_i\, \sigma_i ,
\end{equation}
where $\{\sigma\}_g$ labels all the spins and $E^J_g (\{\sigma\}_g)$
represents the energy function of tile (\textit{i.e.}, gate) $g$.

This Hamiltonian is the starting point for our mapping of universal
classical computation into the Chimera architecture of the D-Wave
machine, one of the important results of the paper, which we discuss
in detail below. In order to anticipate the fact that quantum rather
than classical thermal annealing may be a more effective way of
reaching the ground state and therefore the solution of these
computational problems, we add a transverse magnetic field $\Gamma$ to
Eq. (\ref{eq:E_C}) to obtain the quantum Hamiltonian
\begin{eqnarray}
\label{eq:E_QA2}
\hat H &=& \sum_{g} E^J_g (\{\hat\sigma^z\}_g) - K \sum_{\langle
  i,j\rangle} {{\hat\sigma^z}_i}\, {{\hat\sigma^z}_j} -
\!\!\!\sum_{i\in \text{boundary}} h_i\, \hat\sigma^z_i \nonumber \\ &
& +\ \Gamma \sum_i \hat{\sigma}_i^x .
\end{eqnarray}
However, we find it more expedient and intuitive to work directly with
tiles which satisfy the logic gate constraint exactly. We thus proceed
by projecting the system onto the manifold of states where all local
gate constraints are satisfied by working in the limit in which both
$h$ and $J$ are very large; and we imagine varying $K$ and $\Gamma$,
with $K,\Gamma \ll J,h$ \cite{castelnovo2006}. This limit is best
understood if we switch off the coupling between tiles, $K$. Within a
given tile, the configurations that satisfy the logic gate constraints
span the degenerate ground state manifold, while the unsatisfying
configurations have energies of order $J$ and higher. Let
$\{|q_a\rangle\}$, $a=1,\dots,r$ be all the $r$ states spanned by the
spin configurations $|\sigma_1,\dots,\sigma_n\rangle$ that define the
ground state manifold. For two-bit (four-spin) gates we have $r=4$,
while for three-bit (six-spin) gates $r=8$.

As long as $\Gamma\ll J$ we can understand the effect of a transverse
field $\Gamma$ on the $r$ degenerate states by degenerate perturbation
theory. Since for reversible gates maintaining the gate constraints
requires at least two spin flips, the transverse field $\Gamma$
induces an effective, second-order or higher spin-spin interaction on
the ground state manifold of a given tile of order $\Delta =
\Gamma^2/J$ or lower. This discussion leads naturally to the quantum
vertex model presented in Eq.~(\ref{eq:H-vertex}).

Switching on the $K$ coupling penalizes configurations in which the
states of adjacent tiles are incompatible. Thus, in order to satisfy
both intra- and inter-tile constrains that define the computational
process we must reach the limit of $\Delta \ll K\ll J,h$.

\subsection{Thermodynamics of the classical vertex model}
\label{sec:transfermatrix}

We start by considering the partition function of the classical limit
of the Hamiltonian in Eq.~(\ref{eq:H-vertex}), i.e., $\Delta=0$, which
we obtain via a transfer matrix calculation. Consider first a system
with free boundary conditions at both ends. The partition function for
the vertex model can be more easily written using the spin variables
on the links of the lattice. Let $\{\sigma\}_j$ denote the spin states
on a vertical line, which cuts across $3L$ spins (with $L$ the number
of vertices or 3-bit gates in a column). For convenience we shall
utilize the notation $|\{\sigma\}_j\rangle$ for the vectors in this
transfer matrix calculation.  Within this notation, one can write the
partition function as
\begin{widetext}
\begin{eqnarray}
  Z= \sum_{\{\sigma_{1}\},\dots,\{\sigma_{2W}\}}
  \langle\{\sigma_{1}\}|P_1|\{\sigma_{2}\}\rangle
  \langle\{\sigma_{2}\}\}|T|\{\sigma_{3}\}\rangle
  &\times\cdots\times&
  \langle\{\sigma_{2j-1}\}|P_j|\{\sigma_{2j}\}\rangle
  \langle\{\sigma_{2j}\}|T|\{\sigma_{2j+1}\}\rangle \times\cdots
  \nonumber\\ &\cdots\times&
  \langle\{\sigma_{2W-1}\}|P_W|\{\sigma_{2W}\}\rangle \;,
\end{eqnarray}
\end{widetext}
where the matrix $T$ encodes the energy costs for matching spins
across the links, and the matrices $P_j$ encode the computations
performed by one column of gates. The two types of slices are depicted
in Fig.~\ref{fig:ripple-carry}. Notice that the $T$ is the same for
all slices, and its matrix elements are given by
\begin{equation}
  \langle\{\sigma_{2j}\}|T|\{\sigma_{2j+1}\}\rangle =
  \exp\left({\sum_{a=1}^{3L}
    \beta\,K\,\sigma_{2j,a}\,\sigma_{2j+1,a}}\right),
\end{equation}
whereas $\langle\{\sigma_{2j-1}\}|P_j|\{\sigma_{2j}\}\rangle$
represents the matrix element of $P_j$ at the $j$th column and thus
depends on the particular set of gates within that column. However,
all $P_j$ are permutation matrices since all gates are
reversible. This fact is essential because it allows us to compute the
partition function exactly, irrespective of the circuit.

For the next step notice that the vector $|\Sigma\rangle = \sum_{\{\sigma\}}
|{\{\sigma\}}\rangle$ is an eigenvector of $P_j$ for any operation
$P_j$:
\begin{eqnarray}
  P_j\;|\Sigma\rangle & = & P_j\;\sum_{\{\sigma\}}|{\{\sigma\}}\rangle
  = \sum_{\{\sigma\}} P_j\;|{\{\sigma\}}\rangle =
  \sum_{\{\sigma'\}}|{\{\sigma'\}}\rangle \nonumber \\ & = &
  |\Sigma\rangle \;,
\end{eqnarray}
where we used that we can relabel the states after the
permutation. The vector $|\Sigma\rangle = \sum_{\{\sigma\}}
|{\{\sigma\}}\rangle$ is also an eigenvector of $T$:
\begin{eqnarray}
  T\;|\Sigma\rangle & = & T\;\sum_{\{\sigma\}}|{\{\sigma\}}\rangle=
  \sum_{\{\sigma\},\{\sigma'\}} |\{\sigma'\}\rangle
  \langle\{\sigma'\}|T|\{\sigma\}\rangle \nonumber \\ &=&
  \sum_{\{\sigma'\}} |\{\sigma'\}\rangle \sum_{\sigma}
  \exp\left({\sum_{a=1}^{3L} \beta\,K\,\sigma'_{a}\,\sigma_{a}}\right)
  \nonumber\\ &=& \sum_{\{\sigma'\}} |\{\sigma'\}\rangle
  \;\;\prod_{a=1}^{3L}\;\; \sum_{\sigma_{a}=\pm 1} e^{\;\beta
    K\,\sigma'_{a}\,\sigma_{a}} \nonumber\\ &=& \sum_{\{\sigma'\}}
  |\{\sigma'\}\rangle \; (2\cosh \beta K)^{3L} \nonumber\\ &=& (2\cosh
  \beta K)^{3L}\; |\Sigma\rangle \; \;.
\end{eqnarray}

By collecting all the factors we arrive at the partition function
\begin{eqnarray}
  Z &=& \langle\Sigma|\;P_1\;T\;P_2\;T\dots\;T\;P_{W}\;|\Sigma\rangle
  \nonumber\\ &=& (2\cosh \beta
  K)^{3L(W-1)}\;\langle\Sigma|\Sigma\rangle \;.
\end{eqnarray}
The overlap $\langle\Sigma|\Sigma\rangle=2^{3L}$ reflects the $2^{3L}$
degenerate ground states corresponding to open boundary conditions on
both boundaries.  Had we fixed one of the boundaries to a particular
state $|\{\sigma\}_{\rm fixed}\rangle$ we would have instead obtained
an overlap $\langle \{\sigma\}_{\rm fixed}|\Sigma\rangle=1$. More
generally, in the thermodynamic limit boundaries contribute an
entropic term that counts the number of ground states, but does not
affect the bulk thermodynamics. In particular, the bulk free energy is
that of a paramagnet:
\begin{equation}
  \beta F =-\left[3L(W-1)\right]\; \ln (2\cosh \beta K) \;.
\end{equation}
This also implies that thermodynamics alone, which is independent of
the specific form of the circuit, cannot reveal the complexity of a
ground-state computation, which is reflected in the dynamics of the
system's relaxation into its ground state.

\subsection{Metropolis algorithm for thermal annealing}

The Metropolis simulations are carried out as follows. We work on a
lattice of $L\times W$ vertices, using the Hamiltonian of
Eq.~(\ref{eq:H-vertex}) with $\Delta_{q_s,q_s'}=0$. Periodic boundary
conditions are used in the transverse direction, i.e., the circuit is
laid down on the surface of a tube of length $W$ and circumference
$L$. We consider four types of circuits corresponding to different
concentrations of TOFFOLI gates (the other four types of gates are
assigned equal concentrations): a circuit with only TOFFOLI gates
(100\% concentration), and random circuits with 40\%, 20\%, and 0\%
concentration of TOFFOLI gates.

The first step of the simulation is
to construct a reference state $\{q^{\text{sol}}\}$ that solves the
circuit, by fixing the states $q^{\text{sol}}_s$ for the vertices $s$
at the left boundary, and determining and storing all other states
$q^{\text{sol}}_s$ for vertices $s$ in the rest of the circuit. Next,
we construct three explicit boundary conditions consistent with the
reference state, $\{q^{\text{sol}}\}$, that will serve as the input
states for our simulations: 1. fixed input, for which we apply pinning
fields at the left boundary that fix the states to match
$q^{\text{sol}}_s$ for the vertices $s$ at the left boundary, and
leave the other boundary free (no pinning field); 2. fixed input and
output, for which we pin all vertices $s$ on the left and right
boundaries to those defined by $q^{\text{sol}}_s$; and 3. mixed
boundaries, for which we pin $L_{\partial}=L/2+3$ vertices on both the
left and right boundaries to the solution values $q^{\text{sol}}_s$,
but leave all the remaining boundary vertices free. Our computations
proceed in each of these three cases by averaging over 2000
independent random instances of input states for a given circuit with
a fixed concentration of TOFFOLI gates corresponding to different
$\{q^{\text{sol}}\}$ and computing the average order parameter
$\langle m \rangle$ as a function of the relaxation time $\tau$. Here
it is important to stress that the partial specification of boundaries
in the case of mixed boundary conditions generically leads to multiple
solutions which compete in establishing the local configurations of
gates consistent with the global constraints defining the
computational circuit. While we also discuss cases with multiple
solutions and no solution in Results, the focus of
this paper is on problems with a single solution. To ensure a single
solution in the case of mixed boundary conditions we always check that
each of the random instances of the input state for a given circuit
allows for one and only one solution.

\subsection{Mapping onto the Chimera architecture of the D-Wave machine}

Figure \ref{fig:chimera} shows the `flow chart' of embedding a
$4\times 4$ tile lattice into the Chimera graph. The entire tile
lattice is rotated by $45^\circ$ for convenience, and spins living on
the boundary of each tile are shown explicitly. The lattice of tiles
can be further divided into two sublattices labeled by dark and light
grey, for reasons that should become clear shortly.  Now let us first
consider how to encode the ID and SWAP gates represented by a single
square tile into a unit cell. The embedding involves internal
couplings $J$ that enforce the gate constraints, and the ``grout''
couplings $K$ that match adjacent tiles. The Chimera unit cell forms a
complete bipartite graph $K_{4,4}$, as depicted in
Fig.~\ref{fig:chimera}b-(i), with each spin in one column coupled to
all spins in the other, but not to those in their own column
\cite{Choi}. In order to obtain the generic spin couplings to
represent the gates on a single tile, which inevitably involves
couplings between qubits in the same column as well, we use an
additional ferromagnetic coupling to slave the spins in one column to
their nearest neighbors in the other column. Thus, effectively we are
left with four spins that are fully connected. Details are shown in
Fig. \ref{fig:chimera}b-(i).

In order to use the connectivity of the Chimera graph architecture and
couple adjacent tiles properly, it is convenient to explore the
bipartiteness of the square lattice. Let us take one tile from the
rotated tile lattice, and label the four qubits by their locations on
the tile: N (North), S (South), W (West) and E (East), as shown in
Fig. \ref{fig:chimera}b-(ii). The spins in the adjacent tiles are
matched by the ``grout'' coupling $K$. Upon a careful inspection of
the resulting tile lattice, we notice that the qubits labeled by N and
W in one tile are always connected respectively to qubits S and E in
its neighbor. Therefore, once we fix the embedding of one sublattice
in the unit cell, the embedding of the other sublattice must be
different, because qubits in one unit cell are only coupled to those
at the same place in the neighboring unit cell. We map the two
sublattices of the tile lattice in the unit cells as illustrated in
Fig. \ref{fig:chimera}b-(ii).

Finally, let us show how to embed the TOFFOLI gate, which corresponds
to a rectangular tile, into the unit cells. The TOFFOLI gate can be
viewed as consisting of two square tiles, thus requiring two unit
cells, and the spins coupled between these two tiles exactly provide
the ancilla bit needed in Hamiltonian of
Eq. (\ref{H-Toffoli}). Similar to the square tiles considered above,
within a unit cell we use additional ferromagnetic couplings to slave
qubits from one column to the other when necessary. The explicit
mapping is shown in Fig. \ref{fig:chimera}b-(iii). As we have already
seen, in order to come up with a proper embedding, one has to
carefully take into account how qubits are coupled to adjacent unit
cells. Putting all of the above ingredients together, we arrive at the
embedding of the entire $4\times4$ tile lattice including TOFFOLI into
the Chimera graph, as shown in Fig. \ref{fig:chimera}a.



\vspace{0.8cm}
{\bf Data availability:}
The data that support the main findings of this study are available from the corresponding author upon request.

{\bf Author contributions:}
All authors contributed to all aspects of this work.

{\bf Competing financial interests:}
The authors declare no competing
financial interests.


\newpage

\begin{thebibliography}{99}
\bibitem{montanari_mezard} M. M\'ezard and A. Montanari, {\it
  Information, Physics, and Computation} (Oxford University, Oxford,
  2009).

\bibitem{MPZ} M. M\'ezard, G. Parisi, and R. Zecchina, {\it Analytic
  and algorithmic solution of random satisfiability problems}, Science
  {\bf 297}, 812-815 (2002).

\bibitem{ganguli} S. Ganguli and H. Sompolinsky, {\it Compressed
  sensing, sparsity, and dimensionality in neuronal information
  processing and data analysis}, Ann. Rev. Neuroscience {\bf 35}, 485-508
  (2012).

\bibitem{pankaj} P. Mehta and D. J. Shwab, {\it An exact mapping
  between the variational renormalization group and deep learning},
  Preprint at https://arxiv.org/abs/1410.3831 (2014).

\bibitem{landauer1} R. Landauer, {\it Irreversibility and heat
  generation in the computing process}, IBM J. Res. Dev. {\bf 5}(3),
  183-191 (1961).

\bibitem{landauer2} R. Landauer, {\it The physical nature of
  information}, Phys. Lett. A {\bf 217}, 188-193 (1996).

\bibitem{bennett} C. H. Bennett, {\it The thermodynamics of
  computation--a review}, Int. J. Theor. Phys. {\bf 21}(12), 905-940
  (1982).

\bibitem{nielsen-chuang} M. A. Nielsen and I.  L. Chuang, {\it Quantum
  Computation and Quantum Information} (Cambridge University Press,
  Cambridge 2000).

\bibitem{review} T. D. Ladd, F. Jelezko, R. Laflamme, Y. Nakamura,
  C. Monroe, and J. L. O'Brien, {\it Quantum computers}, Nature
  (London) {\bf 464}, 45-53 (2010).

\bibitem{feynman} R. P. Feynman, {\it Quantum mechanical computers},
  Found. Phys. {\bf 16}, 507-531 (1986).
  
\bibitem{Biamonte2008} J. D. Biamonte, {\it Nonperturbative k-body to
  two-body commuting conversion Hamiltonians and embedding problem
  instances into Ising spins}, Phys. Rev. A {\bf 77}, 052331 (2008).

\bibitem{crosson2010} I. J. Crosson, D. Bacon, and K. R. Brown, {\it
  Making classical ground-state spin computing fault-tolerant},
  Phys. Rev. E {\bf 82}, 031106 (2010).

\bibitem{Biamonte2012} D. Whitfield, M. Faccin and J. D. Biamonte,
  {\it Ground-state spin logic}, Eur. Phys. Lett. {\bf 99}, 57004
  (2012).

\bibitem{Ricci-Tersenghi} F. Ricci-Tersenghi, {\it Being glassy
  without being hard to solve}, Science {\bf 330}, 1639-1640 (2010).

\bibitem{Valiant-Vazirani} L. G. Valiant, and V. V. Vazirani, {\it NP
  is as easy as detecting unique solutions}, Theor. Comp. Sci. {\bf
  47}: 85-93 (1986).

\bibitem{gavey-johnson} M. R. Gavey and D. S. Johnson, {\it Computers
  and Intractability: A Guide to the Theory of NP-Completeness}
  (Freeman, New York, 1979).

\bibitem{papadimitriou} C. H. Papadimitriou and K. Steiglitz, {\it
  Combinatorial Optimization -- Algorithms and Complexity} (Dover,
  Mineola, NY, 1998).

\bibitem{arora-barak} S. Arora and B. Barak, {\it Computational
  Complexity: a Modern Approach} (Cambridge University, New York,
  2009).
  
\bibitem{SA} S. Kirkpatrick, C. D. Gelatt Jr., and M. P. Vecchi, {\it
  Optimization by simulated annealing}, Science {\bf 220}, 671-680 (1983).
  
\bibitem{apolloni1989} B. Apolloni, C. Carvalho, and D. de Falco, {\it
  Quantum stochastic optimization}, Stochastic Process Appl. {\bf 33},
  233-244 (1989).

\bibitem{finnila1990} A. Finnila, M. Gomez, C. Sebenik, C. Stenson,
  and J. Doll, {\it Quantum annealing: a new method for minimizing
    multidimensional functions}, Chem. Phys. Lett. {\bf 219}, 343-348
  (1990).

\bibitem{kadowaki1998} T. Kadowaki and H. Nishimori, {\it Quantum
  annealing in the transverse Ising model}, Phys. Rev. E {\bf 58},
  5355-5363 (1998).
  
\bibitem{farhi2001} E. Farhi, J. Goldstone, S. Gutmann, J. Laplan,
  A. Lundgren, and D. Preda, {\it A quantum adiabatic evolution
    algorithm applied to random instances of an NP-complete problem},
  Science {\bf 292}, 472-475 (2001).

\bibitem{vedral} V. Vedral, A. Barenco, and A. Ekert, {\it Quantum
  networks for elementary arithmetic operations}, Phys. Rev. A {\bf
  54}, 147-153 (1996).

\bibitem{castelnovo2006} C. Castelnovo, C. Chamon, C. Mudry, and
  P. Pujol, {\it High-temperature criticality in strongly constrained
    quantum systems}, Phys. Rev. B {\bf 73}, 144411 (2006).
    
\bibitem{Liu-Polkovnikov-Sandvik} C.-W. Liu, A. Polkovnikov, and
  A. W. Sandvik, {\it Dynamic scaling at classical phase transitions
    approached through nonequilibrium quenching}, Phys. Rev. B {\bf
    89}, 054307 (2014).

\bibitem{kibble} T. W. B. Kibble, {\it Topology of cosmic domains and
  strings}, J. Phys. A: Math. Gen. {\bf 9}, 1378-1398 (1976).

\bibitem{zurek} W. H. Zurek, {\it Cosmological experiments in
  superfluid helium?}, Nature (London) {\bf 317}, 505-508 (1985).
  
\bibitem{Rubin} S. Rubin, N. Xu, and A. W. Sandvik, {\it Dual time
  scales in simulated annealing of a two-dimensional Ising spin
  glass}, Preprint at https://arxiv.org/abs/1609.09024 (2016).

\bibitem{Krapivsky} P. L. Krapivsky, {\it Slow Cooling of an
  Ising Ferromagnet}, J. Stat. Mech. P02014 (2010)

\bibitem{Choi} V. Choi, {\it Minor-embedding in adiabatic quantum
  computation: II. Minor-universal graph design}, Q. Inf.
  Process. {\bf 10}, 343-353 (2011).
  
\bibitem{Newman-Moore1999} M. E. J. Newman, and C. Moore, {\it Glassy
  dynamics and aging in an exactly solvable spin model}, Phys. Rev. E
  {\bf 60}, 5068-5072 (1999).

\bibitem{Garrahan-Newman2000} J. P. Garrahan and M. E. J. Newman, {\it
  Glassiness and constrained dynamics of a short-range nondisordered
  spin model}, Phys. Rev. E {\bf 62}, 7670-7678 (2000).

\bibitem{Ritort-Solich2003} For a review, see F. Ritort and
  P. Sollich, {\it Glassy dynamics of kinetically constraint models},
  Adv. Phys. {\bf 52}, 219-342 (2003).


\end{thebibliography}

\newpage
\onecolumngrid

\title{Supplementary Information \\
  Quantum Vertex Model for Reversible
  Classical Computing}
\date{}

\maketitle

\section*{Supplementary notes}

\subsection*{Supplementary note 1: Building TOFFOLI gates with one-and two-body interactions}

In this section, we present the table with all possible
configurations of a TOFFOLI gate formulated in a rectangular tile, and
their corresponding energies given by
%
\begin{eqnarray}
E_{{\rm TOFFOLI}}(\sigma_a, \sigma_b, \sigma_c; \sigma_{a'},
\sigma_{b'}, \sigma_{d}; \sigma_S) & = & -
J(\sigma_a\sigma_{a'}+\sigma_b\sigma_{b'}) + J(\sigma_a -3\sigma_b -
2\sigma_c +2\sigma_d +4\sigma_S) \nonumber \\ & &
+\ J(-3\sigma_a\sigma_b - 2\sigma_a\sigma_c + 4\sigma_b\sigma_c +
2\sigma_a\sigma_d-4\sigma_b\sigma_d - 4 \sigma_c\sigma_d \nonumber
\\ & & \;\;\;\;\;\;
+\ 4\sigma_a\sigma_S - 8\sigma_b\sigma_S - 6\sigma_c\sigma_S +
6\sigma_d\sigma_S).
\label{H-Toffoli}
\end{eqnarray}
%
 From
which it is clear that the states which satisfy the gate constraint
all stay in the ground state manifold of the Hamiltonian.

The TOFFOLI gate takes a three-bit input sate $(a, b, c)$ into $(a, b,
ab\oplus c)$. The copying of the first two bits is trivially enforced
by a ferromagnetic coupling: $-J(\sigma_a\sigma_{a'} +
\sigma_b\sigma_{b'})$, so we shall only list the part that enforces
the third output bit: $d=ab\oplus c$ in the table below. As explained
in Methods, in order to achieve that with at most
two-body interactions, one needs an ancilla bit that we call $S$.

In Supplementary Table \ref{table:1}, the first eight states separated by double
lines span the ground state manifold of Hamiltonian (\ref{H-Toffoli}),
and one can readily check that they indeed satisfy the gate constraint
imposed by the TOFFOLI gate. All other unsatisfying states have energy
scales higher by multiple of $J$. Therefore in the limit where $J$ is
much larger than any other energy scale in the system ($K, \Gamma$ and
$T$), one essentially projects out the ground state manifold. Notice
that there are cases when $(a, b, c, d)$ satisfies the gate
constraint, but the ground state only picks one value of $S$. For
example, the state $(a, b, c, d)=(0, 0, 1, 1)$ satisfies the gate
constraint, but the ground state manifold only picks $S=0$, and $S=1$
has a higher energy. This does not cause a problem because the ancilla
bit does not enter the computation, and its value does not really
matter.

\subsection*{Supplementary note 2: Nearest neighbor vertex couplings}

The $K^{g_s g_{s'}}_{q_s,q_{s'}}$ couplings encode the energy cost for
mismatched nearest-neighbor vertices. Here we give an example of how
these matrix elements are constructed. Consider two adjacent vertices
at $s$ and $s'$ that enforce the TOFFOLI gate: $g_s=g_{s'}=$TOFFOLI
(or T for short). The matrix elements of $K^{T,T}_{q_s,q_{s'}}$
basically count the number of bits that are mismatched between the
output state $G(q_s)$ of one gate, where $G$ is the gate function, and
the input state $q_{s'}$ of the other gate, for
$q_{s,s'}=0,1,\dots,7$.

When the two sites $s$ and $s'$ share a single bond, the $8\times 8$
matrix $K^{T,T}_{q_s,q_{s'}}$ (with $q_{s,s'}=0,1,\dots,7$) is given
by
\begin{equation}
  K^{T,T}
  =
  \begin{bmatrix}
    0&K&0&K&0&K&0&K\\
    0&K&0&K&0&K&0&K\\
    0&K&0&K&0&K&0&K\\
    0&K&0&K&0&K&0&K\\
    K&0&K&0&K&0&K&0\\
    K&0&K&0&K&0&K&0\\
    K&0&K&0&K&0&K&0\\
    K&0&K&0&K&0&K&0
  \end{bmatrix}
  \;,
\end{equation}
and when the sites share a double bond, the $8\times 8$ matrix is
given by
\begin{equation}
  K^{T,T}
  =
  \begin{bmatrix}
    0&0&K&K&K&K&2K&2K\\
    K&K&0&0&2K&2K&K&K\\
    K&K&2K&2K&0&0&K&K\\
    2K&2K&K&K&K&K&0&0\\
    0&0&K&K&K&K&2K&2K\\
    K&K&0&0&2K&2K&K&K\\
    2K&2K&K&K&K&K&0&0\\
    K&K&2K&2K&0&0&K&K
  \end{bmatrix}
  \;,
\end{equation}
where $K$ is the ferromagnetic energy scale that penalizes
configurations in which the states of adjacent vertices are
incompatible.


\subsection*{Supplementary note 3: Bounds on number of replicas needed for learning algorithm}

Below we estimate the number of replicas, $N_R$ required to achieve
the correct assignment of gates in the "annealing with learning"
algorithm described in Results with accuracy
$\epsilon$. Assume that the probability of correct assignment of a
gate within each replica is $p$. The probability that $k$ of the $N_R$
replicas assign the wrong identity to a given gate is given by the
binomial distribution:
\begin{equation}
\label{eq:binomial}
P(N_R,k;p)  =  \frac{N_R!}{(N_R -k)! k!} p^{N_R - k} (1-p)^k.
\end{equation}
The probability that a fraction greater than $\alpha$ of the $N_R$
replicas ($N_R \gg1$) assign the wrong identity to a particular gate
can then be written as:
\begin{eqnarray}
 P_w (N_R, \alpha;p) &=& \sum_{k=\alpha N_R}^{N_R} \frac{N_R!}{(N_R
   -k)! k!} p^{N_R - k} (1-p)^k\nonumber\\ &<&
 \frac{N_R!}{(\frac{N_R}{2}!)^2}~\sum_{k=\alpha N_R}^{N_R}p^{N_R - k}
 (1-p)^k \\ &\approx & ~\frac{[{2p^{1-\alpha}
       (1-p)^\alpha}]^{N_R}}{\frac{(2p-1)}{p}}\nonumber.
 \end{eqnarray}
It then follows that, for fixed $p>1/2$, $N_R$ and $\alpha$, the
probability that a correct assignment is made to all $LW$ gates by a
fraction of the replicas greater than $\alpha$ is $1-\epsilon$ is
given by:
\begin{eqnarray}
P_c (N_R, \alpha;p) & = & [1-P_w (N_R, \alpha;p)]^{LW} \nonumber \\
&=& 1-\epsilon \\
&>& 1 -  LW ~\frac{[{2p^{1-\alpha} (1-p)^\alpha}]^{N_R}}{\frac{(2p-1)}{p}}
 \nonumber ,
\end{eqnarray}
which implies that the number of replicas, $N_{R\epsilon}$, needed to
ensure an error rate smaller than $\epsilon$ is given by:
\begin{equation}
N_{R\epsilon} = \frac{{\rm ln} \left[ \frac{2p-1}{p}
    \frac{\epsilon}{LW} \right]} {{\rm ln} \left[
    2p^{1-\alpha}(1-p)^{\alpha}\right]} .
\end{equation}
We note that the above argument assumes that the states of the gates
are uncorrelated; for a given $N_R$ the correlations built into the
vertex model should lead to a lower error rate than the estimate given
here.

\begin{table}[H]
\centering
\begin{tabular}{c c c c| c | c c c c | c | c}
\hline
\hline
$a$ & $b$ & $c$ & $d$ & $S$ & $\sigma_a$ & $\sigma_b$ & $\sigma_c$ & $\sigma_d$ & $\sigma_S$ & $E$ \\
\hline
$0$ & $0$ & $0$ & $0$ & $0$ & $-$ & $-$ & $-$ & $-$ & $-$ & $-13J$  \\
\hline
$0$ & $0$ & $1$ & $1$ & $0$ & $-$ & $-$ & $+$ & $+$ & $-$ & $-13J$  \\
\hline
$0$ & $1$ & $0$ & $0$ & $1$ & $-$ & $+$ & $-$ & $-$ & $+$ & $-13J$   \\
\hline
$0$ & $1$ & $1$ & $1$ & $1$ & $-$ & $+$ & $+$ & $+$ & $+$ & $-13J$   \\
\hline
$1$ & $0$ & $0$ & $0$ & $0$ & $+$ & $-$ & $-$ & $-$ & $-$ & $-13J$  \\
\hline
$1$ & $0$ & $1$ & $1$ & $0$ & $+$ & $-$ & $+$ & $+$ & $-$ & $-13J$  \\
\hline
$1$ & $1$ & $0$ & $1$ & $0$ & $+$ & $+$ & $-$ & $+$ & $-$ & $-13J$ \\
\hline
$1$ & $1$ & $1$ & $0$ & $1$ & $+$ & $+$ & $+$ & $-$ & $+$ & $-13J$ \\
\hline
\hline
$0$ & $0$ & $0$ & $1$ & $0$ & $-$ & $-$ & $-$ & $+$ & $-$ & $-9J$ \\
\hline
$0$ & $0$ & $1$ & $0$ & $1$ & $-$ & $-$ & $+$ & $-$ & $+$ & $-9J$  \\
\hline
$0$ & $1$ & $0$ & $1$ & $0$ & $-$ & $+$ & $-$ & $+$ & $-$ & $-9J$  \\
\hline
$0$ & $1$ & $1$ & $0$ & $1$ & $-$ & $+$ & $+$ & $-$ & $+$ & $-9J$  \\
\hline
$1$ & $0$ & $1$ & $0$ & $0$ & $+$ & $-$ & $+$ & $-$ & $-$ & $-9J$ \\
\hline
$1$ & $1$ & $0$ & $0$ & $0$ & $+$ & $+$ & $-$ & $-$ & $-$ & $-9J$ \\
\hline
$1$ & $1$ & $0$ & $0$ & $1$ & $+$ & $+$ & $-$ & $-$ & $+$ & $-9J$  \\
\hline
$1$ & $1$ & $1$ & $1$ & $0$ & $+$ & $+$ & $+$ & $+$ & $-$ & $-9J$   \\
\hline
$1$ & $1$ & $1$ & $1$ & $1$ & $+$ & $+$ & $+$ & $+$ & $+$ & $-9J$   \\
\hline
$0$ & $0$ & $1$ & $0$ & $0$ & $-$ & $-$ & $+$ & $-$ & $-$ & $-J$  \\
\hline
$0$ & $1$ & $0$ & $1$ & $1$ & $-$ & $+$ & $-$ & $+$ & $+$ & $-J$  \\
\hline
$1$ & $0$ & $0$ & $1$ & $0$ & $+$ & $-$ & $-$ & $+$ & $-$ & $-J$  \\
\hline
$1$ & $0$ & $1$ & $0$ & $1$ & $+$ & $-$ & $+$ & $-$ & $+$ & $-J$  \\
\hline
$0$ & $0$ & $0$ & $0$ & $1$ & $-$ & $-$ & $-$ & $-$ & $+$ & $3J$  \\
\hline
$0$ & $0$ & $1$ & $1$ & $1$ & $-$ & $-$ & $+$ & $+$ & $+$ & $3J$   \\
\hline
$0$ & $1$ & $0$ & $0$ & $0$ & $-$ & $+$ & $-$ & $-$ & $-$ & $3J$  \\
\hline
$0$ & $1$ & $1$ & $1$ & $0$ & $-$ & $+$ & $+$ & $+$ & $-$ & $3J$   \\
\hline
$1$ & $1$ & $0$ & $1$ & $1$ & $+$ & $+$ & $-$ & $+$ & $+$ & $11J$  \\
\hline
$1$ & $1$ & $1$ & $0$ & $0$ & $+$ & $+$ & $+$ & $-$ & $-$ & $11J$  \\
\hline
$1$ & $0$ & $0$ & $0$ & $1$ & $+$ & $-$ & $-$ & $-$ & $+$ & $19J$  \\
\hline
$1$ & $0$ & $1$ & $1$ & $1$ & $+$ & $-$ & $+$ & $+$ & $+$ & $19J$  \\
\hline
$0$ & $0$ & $0$ & $1$ & $1$ & $-$ & $-$ & $-$ & $+$ & $+$ & $31J$  \\
\hline
$0$ & $1$ & $1$ & $0$ & $0$ & $-$ & $+$ & $+$ & $-$ & $-$ & $31J$  \\
\hline
$1$ & $0$ & $0$ & $1$ & $1$ & $+$ & $-$ & $-$ & $+$ & $+$ & $55J$  \\
\hline
\hline
\end{tabular}
\caption{All possible configurations of a TOFFOLI gate formulated in a
  rectangular tile with an ancilla $S$, and their corresponding
  energies given by Eq. (\ref{H-Toffoli}). The first eight states span
  the ground state manifold, and they satisfy the gate constraint.}
\label{table:1}
\end{table}


\end{document}